\documentclass[12pt]{article}
\usepackage{soul}
\usepackage{color}
\usepackage{xcolor}

\usepackage{moreverb} 

\immediate\write18{texcount -inc -incbib 
-sum main.tex > /tmp/wordcount.tex}

\usepackage[title]{appendix}

\usepackage{newtxtext,newtxmath}

\usepackage{graphicx}

\usepackage[letterpaper,margin=1in]{geometry}

\usepackage{color,soul}

\renewenvironment{abstract}
	{\quotation}
	{\endquotation}

\date{}

\renewcommand\refname{References and Notes}

\makeatletter
\renewcommand{\fnum@figure}{\textbf{Figure \thefigure}}
\renewcommand{\fnum@table}{\textbf{Table \thetable}}
\makeatother

\usepackage{scicite}

\usepackage{url}

\newcommand{\bfd}{{\bf d}}
\newcommand{\bfg}{{\bf g}}

\newcommand{\bfn}{{\bf n}}

\newcommand{\bfx}{{\bf x}}
\newcommand{\bfy}{{\bf y}}

\newcommand{\bfJ}{{\bf J}}
\newcommand{\bfT}{{\bf T}}

\newcommand{\calC}{{\cal C}}
\newcommand{\calD}{{\cal D}}

\newcommand{\calG}{{\cal G}}
\newcommand{\calL}{{\cal L}}
\newcommand{\calM}{{\cal M}}

\newcommand{\calP}{{\cal P}}
\newcommand{\calR}{{\cal R}}
\newcommand{\calS}{{\cal S}}

\newcommand{\bfh}{{\bf h}}
\newcommand{\bfm}{{\bf m}}

\newcommand{\bfs}{{\bf s}}

\newcommand{\bfv}{{\bf v}}

\newcommand{\bfG}{{\bf G}}

\newcommand{\calB}{{\cal B}}

\newcommand{\calX}{{\cal X}}

\newcommand{\begeq}{\begin{equation}}
\newcommand{\begeqarray}{\begin{eqnarray}}
\newcommand{\sign}{\text{sign}}

\def\scititle{
	Inconsistency and Acausality in Bayesian Inference for Physical Problems
}
\title{\bfseries \boldmath \scititle}

\author{
	Klaus~Mosegaard$^{1\ast}$,\and
	Andrew~Curtis$^{2}$\and
	\small$^{1}$Niels Bohr Institute, University of Copenhagen, Copenhagen, 2100, Denmark.\and
	\small$^{2}$ University of Edinburgh, Edinburgh, EH9 3JW, United Kingdom.\and
	\small$^\ast$Corresponding author. email: mosegaard@nbi.ku.dk
}

\begin{document}
\maketitle
 
\begin{abstract}
Bayesian inference is used to estimate continuous parameter values given measured data in many fields of science. The method relies on conditional probability densities to describe information about both data and parameters, yet the notion of conditional densities is inadmissible: probabilities of the same physical event, computed from conditional densities under different parameterizations, may be inconsistent. We show that this inconsistency, together with acausality in hierarchical methods, invalidate a variety of commonly applied Bayesian methods when applied to problems in the physical world, including trans-dimensional inference, general Bayesian dimensionality reduction methods, and hierarchical and empirical Bayes. Models in parameter spaces of different dimensionalities cannot be compared – invalidating the concept of natural parsimony, the probabilistic counterpart to Occams Razor. Bayes theorem itself is inadmissible in non-discrete spaces, and Bayesian inference applied to parameters that characterize physical properties requires reformulation.
\end{abstract}

\section{Introduction}
Bayes Theorem (or Bayes Rule, or Bayes Formula) \cite{Bayes1763,Laplace1774} is a trivial consequence of the definition of conditional probability, expressing how probabilities depend on knowledge about possible outcomes. This has turned out to be immensely useful in scientific inference, where information from independent sources are combined to obtain probabilistic results. For this reason, Bayes Rule is often seen as a model for learning \cite{Bishop2011}: Initial (prior) information is acquired and quantified through a (prior) probability distribution, and subsequently new information (data) is introduced, adding to the prior information. Our resulting, total information can be expressed through Bayes Formula, providing a (posterior) probability distribution, constrained by both the prior and data information. If we only consider discrete outcome spaces (discrete probability distributions), the above inference/learning scheme is rigorous and unproblematic. However, when applied to inference problems involving physical quantities, where probability densities over multi-dimensional manifolds are sought, a serious flaw in the formalism appears. It was discovered already by Bertrand (1889), and later confirmed by Borel (1909) and Kolmogorov (1933), that conditional probability densities over multi-dimensional manifolds are, in fact, inadmissible. In practice, this means that conditional probability densities, and hence Bayesian inference, are not invariant under coordinate/parameter changes. This is known as the Borel-Kolmogorov paradox (herein referred to as the {\it BK inconsistency}).

Seen from a purely statistical viewpoint, this may seem unimportant as long as we stick to the same set of parameters, and this may be the reason why the problem has been, and remains, largely neglected.
In physical sciences, however, invariance of any statement about nature under reparameterization is essential: The physical properties of reality, and our information about it, will not depend on its numerical representation 
\cite{Einstein1916,Jeffreys1946,Norton1993,Thorne2017}. If we accept that any parameter of a physical system and its probability distribution represents a (possibly observer-dependent) physical property, the {\em principle of covariance} imposes invariance under reparameterization: mathematically, its expression in different coordinates/parameters must change only according to a Jacobian transformation. Bayesian inversion schemes that do not satisfy this principle may lead to physically unacceptable solutions. 

In this paper we first expose the BK-inconsistency by showing that currently accepted solutions are not only philosophically unacceptable, they also lead to contradictory results. We proceed by showing that the BK-inconsistency has serious consequences for commonly used Bayesian methods. First, we demonstrate that hierarchical/empirical Bayes methods lead to conflicting probabilistic conclusions when considering different parameterizations of the same data. Then we turn to trans-dimensional inversion procedures and probabilistic analysis of sparse models, and show that the optimal number of model parameters also depends on the data parameterization.
Key is that all such methods are based on the concept of evidence, which we show is inadmissible in a physical inversion context.

Another essential pillar in physical and social sciences is causality -- that effect comes after cause. Similarly to parameterization invariance, causality plays a less clear role in statistics, probably because weaker notions such as `correlation' and `conditioning' dominate in the literature. The dividing line between cause and effect, {\em prior} and {\em posterior}, is defined as the point at which the physical (forward) relations between the unknowns and new data are introduced \cite{Bernardo2000,Jaynes2003}; we shall show that hierarchical/empirical Bayes represents a breach of logical causality, producing posterior results which masquerade as prior information, with serious consequences for inferences concerning the physical world. This problem will remain even if the BK-inconsistency is avoided in a future reformulation of Bayesian selection methods.

The essence of this paper is to demonstrate mathematical inconsistency and logical acausality of commonly used Bayesian methods. The demonstrations are based on {\em reductio ad absurdum}, that is, analytical examples where the methods in question, applied to physical inference problems, are shown to be mathematically self-contradictory or in other ways unacceptable. The results expose deep philosophical arguments against Bayesian model selection methods when applied to natural physical problems, to which a practical response is needed urgently.

\subsubsection*{The basic assumptions}
We  analyze the extent to which Bayesian inference used in Bayesian inversion (solving physical inverse problems using Bayes Theorem) satisfies $3$ fundamental principles assumed in modern physical science:
\renewcommand{\theenumi}{\Alph{enumi}}
\begin{enumerate}
\item 
Uncertainty is always represented by probability distributions, describing the (incomplete) state of information that is available to an observer/analyst \cite{deFinetti19701974,deFinetti19701975,Tarantola82a,Jaynes1984,Bernardo2000,Malinverno2004,Shiffrin2008}.
\label{item:Info-level}
\item
Any observable or non-observable property of a physical system under consideration, and its corresponding probability distribution, must be invariant under reparameterization \cite{Einstein1916,Jeffreys1946,Norton1993,Thorne2017}.
\label{item:Consistency}
\item
A priori information about observables (``data'') and non-observables (``unknown parameters'') changes to new and updated\ {\em a posteriori} information at the moment where we introduce the functional relation between them \cite{Bunge1959,Green2003,Jaynes2003}. In the words of Bernardo and Smith (2000): The prior probability distribution represents beliefs prior to conditioning on data.
\label{item:Causality}
\end{enumerate} 
Assumption \ref{item:Info-level} means that prior probability distributions over observable or non-observable parameters (often referred to simply as the {\em prior}) reflect the initial state of information of the analyst (possibly based on previous data analysis). 
Assumption \ref{item:Consistency} is one of the most important assumptions in physics; it ensures that analysts with the same information, but using different parameterizations, will arrive at the same result.
Assumption \ref{item:Causality} is a statement of logical causality, the  principle that cause comes before effect. This is one of the most crucial principles in the logical description of all physical systems.


\subsection*{Bayes Theorem and Inverse Problems}
\label{sec: BayesIntro}

Let us first describe the classical process of so-called {\em Bayesian inversion}. The method uses Bayes theorem, which in our context is expressed as
\begin{equation}
p_{m|d} (\bfm | \bfd) = \frac{p_{d|m} (\bfd | \bfm)p_m(\bfm)}{p_d(\bfd)}
\label{eq:Bayes-precise}
\end{equation}
where $\bfd$ specifies the data values (observations) and $\bfm$ contains model parameters with unknown values. Terms $p_m$, $p_d$, $p_{m|d}$ and $p_{d|m}$ denote specific marginals and conditionals of the joint distribution $p_{d,m}$.
Bayes theorem is a trivial consequence of the definition of {\em conditional probability}, e.g., $p(\bfd | \bfm) \equiv p (\bfd,\bfm) / p(\bfm)$, for $p(\bfm) \neq 0$.

$p_d(\bfd)$ is the {\em prior} distribution of the noisy, observed data, and $p_m(\bfm)$ is the {\em prior} distribution of the unobservable model parameters $\bfm$, independently of the physical relation between $\bfd$ and $\bfm$. The conditional $p_{d|m} (\bfd | \bfm)$ is termed the {\em likelihood} function, and is related to the {\em forward relation} $\bfd = g(\bfm)$ provided by mathematical physics or statistical correlations. $p_{m|d} (\bfm | \bfd)$ is the {\em posterior probability density}, the result of combining the information in $p_m(\bfm)$ with $p_d(\bfd)$, using the forward relation.

\bigskip\noindent
There are two equivalent formulations of Bayes Theorem for inversion:

\subsubsection*{Formulation 1}
Following the most widely used formulation, let us express the observations thro\-ugh the random variable $\hat{\bfd} = \bfg(\bfm) + \hat{\bfn}$, where $g(\bfm)$ is the forward function giving us the computed data, and $\hat{\bfn}$ is the random variable representing noise or errors incurred when attempting to observe the computed data. Hence, 
\begin{equation}
p_{\hat{d}|m} (\hat{\bfd} | \bfm) = p_{\hat{n}}(\hat{\bfd}-g(\bfm)) 
\end{equation}
where $p_{\hat{n}}$ is the noise distribution, and we obtain
\begin{equation}
p_{m|\hat{d}} (\bfm | \hat{\bfd}) = \frac{p_{\hat{n}}(\hat{\bfd}-g(\bfm))\ p_m(\bfm)}{p_d(\hat{\bfd})}
\end{equation}
For a specific measurement $\bfd_{obs}$ (a realization of $\hat{\bfd}$),
\begin{equation}
p_{m|\hat{d}} (\bfm | \bfd_{obs}) = \frac{p_{\hat{n}}(\bfd_{obs}-g(\bfm))\ p_m(\bfm)}{p_d(\bfd_{obs})}
\label{eq: Bayes-1}
\end{equation}
which is the (conditional) posterior probability density over the model space for the given observations. The distribution $p_{\hat{n}}(\bfd_{obs}-g(\bfm))$ is the likelihood function.

\subsubsection*{Formulation 2}
To streamline our mathematical exposition in this paper we use a second -- equivalent -- formulation \cite{Tarantola82a} of Bayes Theorem which differs from the standard formulation (Formulation 1) in two ways:
\begin{itemize}
    \item We define a random function $\bfd = \bfd_{obs} + \bfn$ whose probability density $p_d(\bfd) = p_n(\bfd - \bfd_{obs})$ is a model of a concrete measurement $\bfd_{obs}$, equipped with error statistics described by the dispersion (e.g., standard deviation) of the noise $\bfn$. In other words, $\bfd$ is a random variable that includes {\em both} the observations and the noise -- which differs conceptually from the definition of deterministic variable $\bfd$ in formulation 1.
    \item The noise $\bfn$ in this formulation is the same as in Formulation 1, except for the sign: $\bfn = -\hat{\bfn}$.
\end{itemize}
 In this formulation, the likelihood function in eq. \ref{eq: Bayes-1} becomes
\begin{equation}
p_{\hat{n}}(\bfd_{obs}-g(\bfm)) = p_n(g(\bfm) - \bfd_{obs}) = p_d(g(\bfm)) 
\label{eq: 2nd likelihood}
\end{equation}
and Bayes' Formula becomes
\begin{equation}
p_{m|\hat{d}} (\bfm | \bfd_{obs}) = \frac{p_d(g(\bfm))\ p_m(\bfm)}{p_d(\bfd_{obs})} 
\label{eq: Bayes-2}
\end{equation}
or, alternatively,
\begin{equation}
p_{m|\hat{d}} (\bfm | \bfd_{obs}) = \frac{p_{d,m}(g(\bfm),\bfm)}{p_d(\bfd_{obs})} 
\label{eq: Bayes-2alt}
\end{equation}

\begin{figure}
\begin{center}
\includegraphics[width=4.3in]{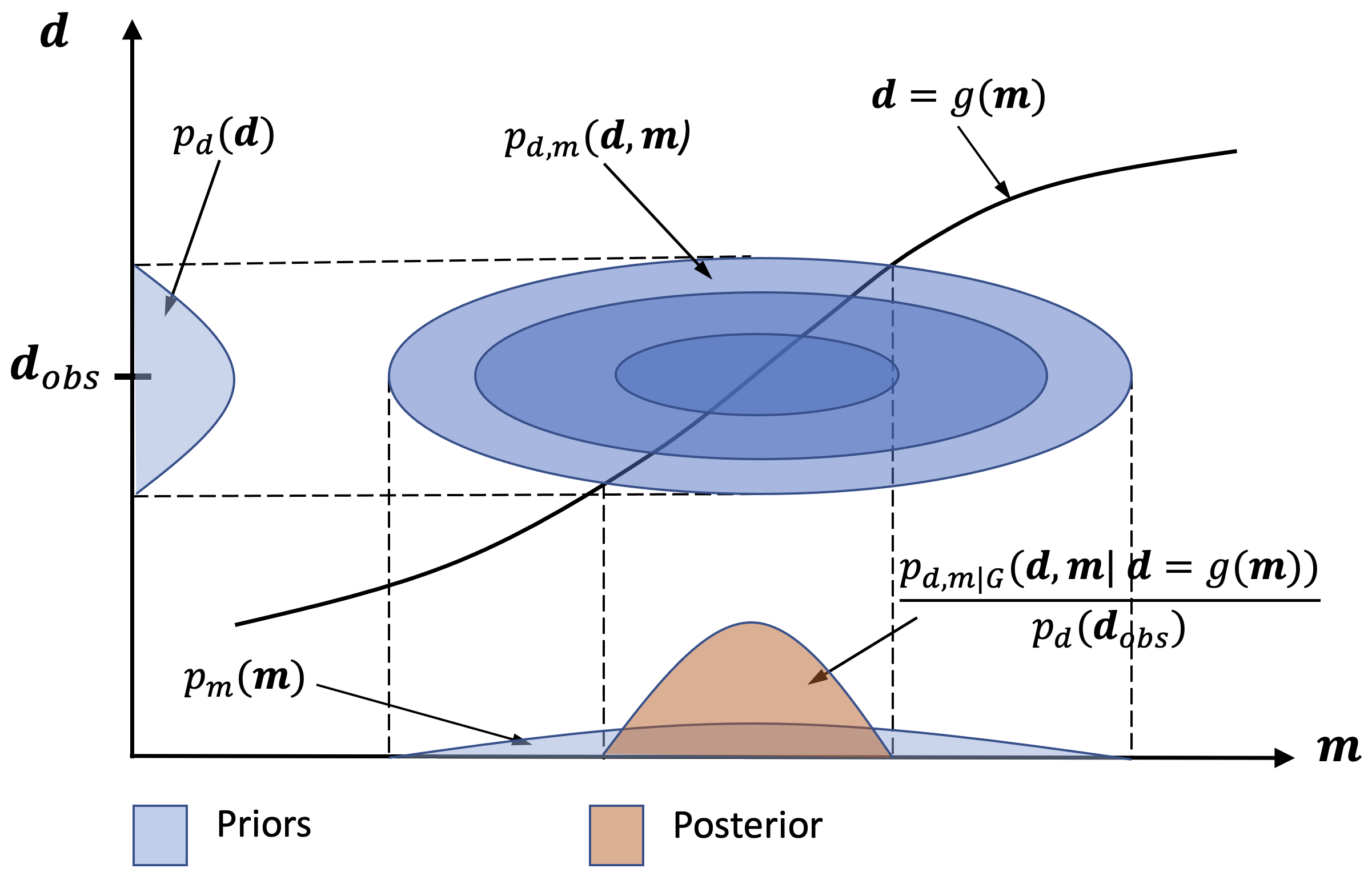}
\caption{\small A graphical illustration of Bayesian inversion, using formulation 2. The joint prior probability density $p_{d,m}$ is the product of the prior probability densities $p_m$ and $p_d$, and $g(\bfm)$ is the forward function. The distribution $p_{d,m|\bfd=g(\bfm)}(\bfd,\bfm|\bfd=g(\bfm))/p_d(\bfd)$ is the posterior, which is often denoted $p_{m|d} (\bfm | \bfd)$ or just $p (\bfm | \bfd)$.}
\label{fig:Bayes}
\end{center}
\end{figure}

In the following, we use this second formulation of Bayesian inversion. A graphical illustration of this scheme is seen in figure \ref{fig:Bayes}. 

\section*{Bayesian Model Selection through Evidences and Bayes Factors}
A Bayesian model selection problem is one where structural or statistical parameters $\bfh$, usually termed {\em hyperparameters}, instead of being given beforehand, are included in the set of unknown variables:
\begin{equation}
\begin{aligned}
p_{m,h|d}(\bfm^{(\bfh)},\bfh|\bfd)) &= \frac{p_{d|m,h}(\bfd|\bfm^{(\bfh)},\bfh)p_{m|h}(\bfm^{(\bfh)}|\bfh)p_h(\bfh)}{p_d(\bfd)} \\
&= \frac{p_{d}(g_h(\bfm^{(\bfh)}))p_{m|h}(\bfm^{(\bfh)}|\bfh)p_k(\bfh)}{p_d(\bfd)}\ .
\end{aligned}
\label{eq: Bayesian-model-select}
\end{equation}
Here, $\bfh$ may include discrete parameters, such as the dimensionality of parameter space as in a later section on {\em trans-dimensional inversion}, or an index referring to different parameterizations with the same dimensionality. Alternatively, $\bfh$ consists of continuous parameters, such as the variances of data uncertainties or physical model priors, as in our discussion of {\em hierarchical Bayes}. The distributions $p_d$, $p_h$, $p_{m|h}$, $p_{m,h|d}$ and $p_{d|m,h}$ are marginals and conditionals derived from the joint prior probability $p_{d,m,h}$. 
We have also assumed that all of the prior probability distributions $p_d$, $p_{m|h}$ and $p_h$ are independent. 

According to Formulation 2 of Bayesian inversion presented above, $p_{d}(g_h(\bfm^{(\bfh)}))$ is the likelihood and $p_{m|h}(\bfm^{(\bfh)}|\bfh)$ is the prior, both now conditioned on $\bfh$. In this context, $p_d(\bfd)$ is termed the {\em total evidence}:
\begin{equation}
p_d(\bfd) = \sum_{\hat{\bfh}} \int p_{d|h}(\bfd | \bfh) p_h(\bfh) d\tilde{\bfh}
\end{equation}
where $\bfh = (\hat{\bfh},\tilde{\bfh})$ may be divided into a discrete part $\hat{\bfh}$ and a continuous part $\tilde{\bfh}$, and where the individual {\em evidences} (or {\em marginal likelihoods}) for each value of $\bfh$ are given by
\begin{equation}
p_{d|h}(\bfd | \bfh) = \int_{\calM} p_d(g_k(\bfm^{(\bfh)}))p_{m|h}(\bfm^{(\bfh)}|\bfh) \ d\bfm^{(\bfh)} \ .
\label{eq:evidence}
\end{equation}
Bayesian model selection is used to compute the posterior probability distribution in the joint $(\bfm^{(\bfh)},\bfh)$-space, allowing comparison of posterior probabilities, not only amongst models with the same hyperparameters, but also across models with different hyperparameters.

In this method it is claimed that the evidence $p_{d|h}(\bfd | \bfh)$ measures how well the hypothesis that our model with hyperparameters $\bfh$ explains the data (under the given prior information). A common measure of how favorable a hypothesis (a value of) $\bfh'$ is, compared to a hypothesis $\bfh''$, is the 
ratio of posterior probabilities of $\bfh$:
\begin{equation}
\frac{p_{h|d}(\bfh' | \bfd)}{p_{h|d}(\bfh'' | \bfd)} = \frac{p_d(\bfd | \bfh')}{p_d(\bfd | \bfh'')} \frac{p_h(\bfh')}{p_h(\bfh'')}\ ,
\end{equation}
where the first ratio on the right-hand-side is the so-called {\em Bayes Factor}, the ratio of the evidences of $\bfh'$ and $\bfh''$. The more this factor increases, the more we prefer $\bfh'$ over $\bfh''$ given the data $\bfd$. Clearly, the evidence plays a key role in Bayesian model selection, since it decides which models are favorable.

\section*{The Borel-Kolmogorov Inconsistency}

We now investigate the BK-inconsistency \cite{Bertrand1889,Borel1909,Kolmogorov1933} and its consequences for Bayesian inversion of physical data in general, and Bayesian model selection in particular.

Given a probability distribution over a space  $ \calX $ represented by
the probability density $ f(\bfx) $, and given a subspace $
\calB $  of  $ \calX $  of lower dimensionality, can we {\em uniquely} infer a probability distribution over $ \calB $,
represented by a probability density $ f(\bfx|\calB) $  (to be
named the conditional probability density `given  $ \calB $')?
According to Borel (1909) and Kolmogorov (1933), the answer to this question is no. In fact, the usual, na\"{i}ve use of conditional probability densities may lead to self-contradicting results, as we demonstrate in the following.

\begin{figure}
\begin{center}
\includegraphics[width=2.6in]{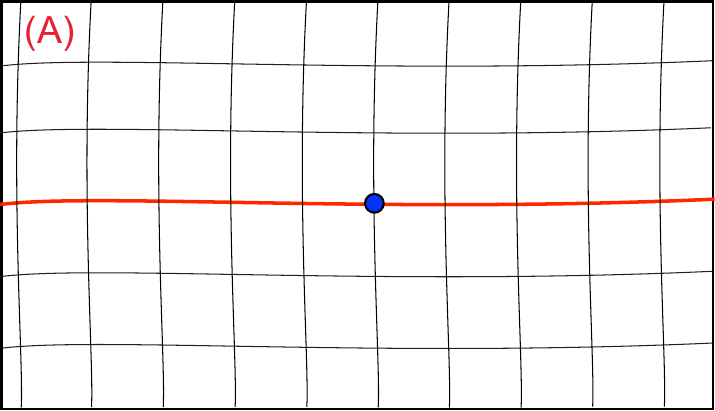}
\quad
\includegraphics[width=2.6in]{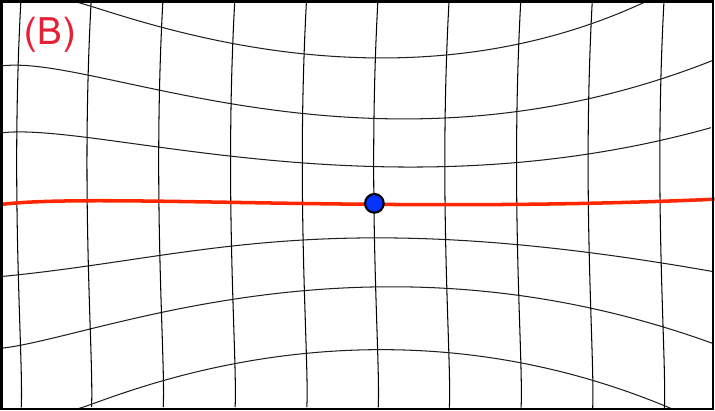}
\caption{\small Two different coordinate grids in the same physical space. A probability distribution assigning equal probabilities to equal volumes is defined over the space. In parameterization (A) the gridline spacing, and hence the probability density, is almost constant. In parameterization (B) the spacing between horizontal gridlines is smaller around the blue mark, and hence the probability density has a minimum in this area. (Modified from Mosegaard and Hansen (2016)).}
\label{fig:Borel-in-2-frames}
\end{center}
\end{figure}

Figure \ref{fig:Borel-in-2-frames} shows two different coordinate systems $A$ and $B$ defined over the same two‐dimensional space. The vertical gridlines and the single red horizontal grid line are identical in systems A and B. The horizontal and vertical coordinate gridlines of system $A$, and the vertical gridlines of
system $B$, are approximately equidistant from their neighbours, but the horizontal gridlines of system $B$ are not. A one‐dimensional subspace (the horizontal red line) located in the middle of the figure coincides with a horizontal gridline. Because the vertical gridlines are the same in system $A$ and $B$, the local coordinate grid in the one‐dimensional subspace is identical in the two systems.

Define a probability distribution over the 2D space assigning equal probabilities to equal physical volumes (areas). Since the gridline spacing in coordinate system $A$ is almost constant, the corresponding probability density in $A$ will be almost constant. In coordinate system $B$, the spacing between horizontal gridlines is smaller in the middle (around the blue mark), and hence the probability density in $B$ has a minimum in this area. This is necessary because, if the coordinates are called $x$ and $y$, the physical volume/area of $[x,x + dx] \times [y,y + dy]$ is smallest in $B$; and considering that $f(x, y) dx dy$ is the probability in $[x,x + dx] \times [y,y + dy]$ for infinitesimal $dx$ and $dy$, the density $f(x, y)$ must be smaller in $B$ to ensure that “equal volumes have equal probabilities” in the physical space.

\begin{figure}
\begin{center}
\includegraphics[width=2.6in]{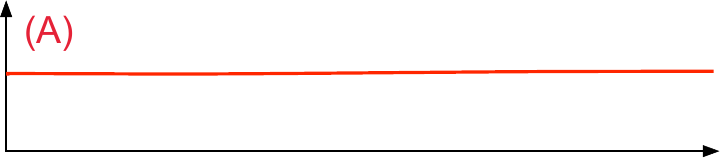}
\quad
\includegraphics[width=2.6in]{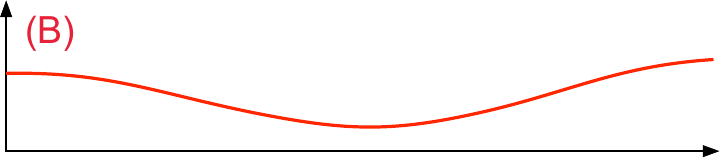}
\caption{\small Conditional probability densities on the (red) 1D subspaces in Figure \ref{fig:Borel-in-2-frames}. Left: The conditional density calculated in system (A). Right: The conditional density calculated in system (B). Note how volume variations between gridlines in the two‐dimensional space in Figure \ref{fig:Borel-in-2-frames} influence the calculated conditionals on the red line, despite the fact that (1) the 2D distributions in both systems assign equal probabilities in each volume, and (2) the grid system on the 1D line is identical in the two systems. This is an example of the Borel-Kolmogorov inconsistency. (Figures modified from Mosegaard and Hansen (2016)).}
\label{fig:Conditionals-in-2-frames}
\end{center}
\end{figure}

However, figure \ref{fig:Conditionals-in-2-frames} shows the conditional probability densities on the one‐dimen\-sional subspace computed in systems $A$ and $B$. Although the same, uniform two‐dimensional probability distribution over physical space is represented in the two coordinate systems, and despite the fact that the local coordinate grid in the one‐dimensional subspace is identical in $A$ and $B$, the two conditional distributions are different. Here we are observing the BK-inconsistency where the probability density, restricted to a fixed subspace, is influenced by a deformation of gridlines, not (only) in the subspace itself but in the surrounding higher‐dimensional space in which it is embedded. In other words, the coordinate transformation that ensures consistency between probability densities in the higher-dimensional space, does not work in the lower-dimensional subspace.

A consequence of the BK-inconsistency, that will be important for the following discussion, is that the {\em integral} of a conditional density will also, in general, change with the parameterization. In fact, as demonstrated in Appendix \ref{Appendix: Non-uniqueness}, it is possible to find a reparameterization  that gives any desired value for the integral.


\subsection*{What happens if we ignore the Borel-Kolmogorov problem?}
We now show that by ignoring the Borel-Kolmogorov problem, two apparently correct inversions based on the same information lead to quite different, contradictory results. 
 
\begin{figure}
\begin{center}
\includegraphics[width=3.5in]{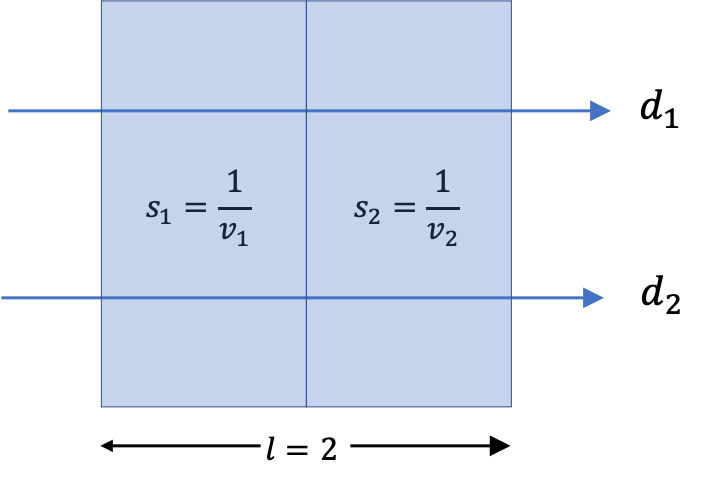}
\caption{\small Simple tomographic example with two model parameters in two different parameterizations (wave speeds $v_i$ or wave slownesses $s_i$) and two traveltime observations.} 
\label{fig:Simple-tomo}
\end{center}
\end{figure}

Consider a simple tomographic example of Bayesian Inversion in two different parameterizations. We consider a case where data $\bfd$ and unknown parameters $\bfm$ are connected through the forward relation
\[
\bfd = g(\bfm)
\]
where $\bfd$ has 2 components (two observations), and $\bfm$ is a model parameter vector with 2 components. Data $\bfd$ contains the travel times along two straight-line rays passing through a medium consisting of 2 homogeneous blocks of width $l=1$ unit, characterized by their wave speeds $v$ or their wave slownesses $s = 1/v$ (figure \ref{fig:Simple-tomo}). 

\begin{figure}
\begin{center}
\includegraphics[width=4.3in]{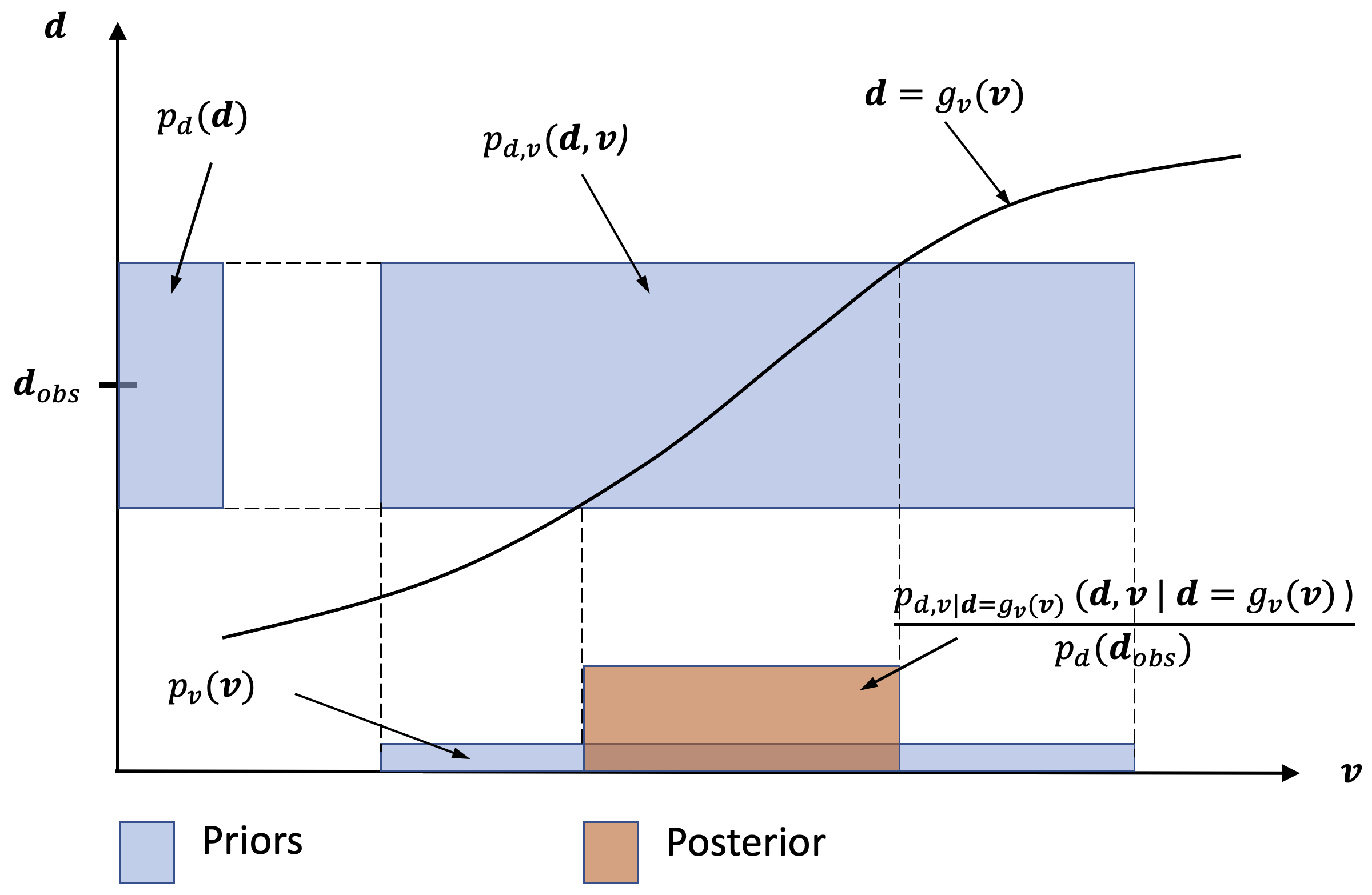}
\caption{\small Bayesian inference of velocity parameters in the special case where the data prior and the model prior are constant in an interval. The joint prior probability density $p_{\bfv,\bfd}$ is the product of the prior probability densities $p_\bfv$ and $p_\bfd$, and $g_v(\bfv)$ is the forward function.}
\label{fig:Bayes-uniform}
\end{center}
\end{figure}

For simplicity we assume that the uncertainty on $\bfd$ has a uniform probability density $p_d$ on a subset of the data space $\calD$, and that the prior on $\bfm$ has a uniform probability density $p_m$ on a subset of the model space $\calM_v$, when using velocity parameters (see figure \ref{fig:Bayes-uniform}).

First, let the unknowns be velocity parameters $\bfv$, with a uniform prior $p_m (\bfv)$ over a subset $\calM_v$ of the model space. In other words, if the forward relation is $\bfd = g_v (\bfv)$, the problem has the simple structure shown in figure \ref{fig:Bayes-uniform}. Under simple conditions outlined in Appendix~\ref{Appendix: Tomo-Ex}, the posterior probability density is: 
\begin{equation}
    p_{v|d} (\bfv | \bfd_{obs})  = 
\begin{cases}
    K' & {\rm for} \ [g_v(\bfv) \in \calD] \wedge [\bfv \in \calM_v]  \\ 
    0              & \text{otherwise}
\end{cases}
\label{eq:Posterior-v_AC}
\end{equation}
where $K'$ is a normalization constant. Thus, the posterior $p_{v|d} (\bfv | \bfd_{obs})$ is homogeneous, since it is equal to the homogeneous, joint prior on $\calM_v \times \calD$, conditioned (evaluated) on the manifold given by $\bfd=g_v (\bfv)$, and normalized so that it integrates to $1$. 

Say we are now interested in the inversion result under the additional constraint that the velocities are the same in the two blocks. From (\ref{eq:Posterior-v_AC}) we compute the {\em conditional} posterior on the 1-D subspace defined by $v_2 = v_1$, as a function of $v_1$:
\begin{equation}
\begin{aligned}
    p_{v_1|d} (v_1 | \bfd_{obs}) 
    &=   
\begin{cases}
    C' & {\rm for} \ [g_v({v_1,v_1}) \in \calD] \wedge [({v_1,v_1}) \in \calM_v]  \\ 
    0              & \text{otherwise}
\end{cases}
\end{aligned}
\label{eq:Cond-Posterior-v_AC}
\end{equation}
where $C'$ is  constant. This density is, of course, constant over an interval.

Let us now see whether this result remains true if we reparameterize from velocities to slownesses: $\bfv = (v_1,v_2)^T \rightarrow \bfs = (s_1,s_2)^T$ through transformation $h$:
\begin{align}
         \begin{pmatrix}
           s_1 \\
           s_2 
         \end{pmatrix}
       = h\begin{pmatrix}
           v_1 \\
           v_2 
         \end{pmatrix}
       = \begin{pmatrix}
           1/v_1 \\
           1/v_2 
         \end{pmatrix}
\label{eq:v-to-s_AC}
\end{align}
By transforming the velocity prior and forward function to this new parametrization, Appendix~\ref{Appendix: Tomo-Ex} shows that under the same conditions we obtain the transformed posterior probability density for $\bfs$,
\begin{align}
    p_{s|d_{obs}} (\bfs | \bfd_{obs}) 
    &\propto 
\begin{cases}
           s_1^{-2} s_2^{-2} & {\rm for} \ [g_s(s_1,s_2) \in \calD] \wedge [(s_1,s_2) \in \calM_s] \\
           0 & \text{otherwise}          
\end{cases}
\label{eq:Posterior-s_AC}
\end{align}
From (\ref{eq:Posterior-s_AC}) we can again compute the posterior, conditioned on  $v_2=v_1$, or  equivalently $s_1=s_2$, and transform the result back to the original velocity parametrization. When formulated as a function of $v_1$, we obtain the 1-D density: 
\begin{equation}
p_{v_1|d} (v_1 | \bfd_{obs})  \propto 
 \begin{cases}
    v_1^2 & {\rm for} \ g_v({v_1,v_1}) \in \calD \wedge ({v_1,v_1}) \in \calM_v  \\ 
    0  & \text{otherwise}
\end{cases}
 \\
\label{eq:Cond-Posterior-v-alt_AC}
\end{equation}
which is in direct contradiction to expression (\ref{eq:Cond-Posterior-v_AC}).

This illustrates the fallacy of using the intuitively straightforward, but incorrect notion of conditional probability density. The problem \cite{Kolmogorov1933} is that our conditional probability is a density on a subspace with zero measure (zero probability) in the original, higher-dimensional space. The Jacobian transformation ensured proper transformation of probabilities from the 2D velocity space to the 2D slowness space -- equations~(\ref{eq:Posterior-v_AC}) and~(\ref{eq:Posterior-s_AC}) describe exactly the same posterior distribution. However, this transformation was locally not the same as the correct transformation between the corresponding 1D subspaces. This implies that one cannot simply assign values from the higher-dimensional density to the lower-dimensional density at co-located points.

Another, practically very important example of the BK-inconsistency is given in Appendix \ref{Appendix: Inconsistent-MAP}. There we expose that the maximum a posteriori (MAP) solution, which is the conditional over a zero-dimensional subspace of parameter space, is also inconsistent \cite{Tarantola1987}.

If we try to remove the BK-inconsistency by defining a lower-dimensional conditional as the limit of a sequence of densities in the original space, we face the problem that there are infinitely many sequences available for this limiting process, leading to different results; indeed, this property fundamentally explains the BK-inconsistency  \cite{Kolmogorov1933,Mosegaard-Tarantola-2002}.


\section*{Inconsistency of Bayesian Model Selection Methods}
\subsection*{Hierarchical Bayes and the BK-inconsistency}

We now turn our attention to the consequences of the BK-inconsistency in the methods of hierarchical and empirical Bayes. Consider an inverse problem where the noise distribution of data, and/or the a priori distribution of model parameters, are unknown. In a standard formulation of Bayesian inversion this would render the problem unsolvable, but the methods of {\em hierarchical Bayes} \cite{Casella1985,Gelman1997}, and its close relative, {\em empirical Bayes} \cite{Good1956,Robbins1956,Carlin2000}, apparently offer a way out: Let $\boldsymbol{\theta}$ be a vector of parameter(s) that define the model prior (e.g., means and standard deviations for a fixed family of distributions), and the data/noise uncertainty (e.g., means and standard deviations of measurements). Hierarchical Bayes now uses equation \ref{eq: Bayesian-model-select} with $\bfh = \boldsymbol{\theta}$.


As we show below, there is a fundamental problem with hierarchical Bayes, rooted in the BK-inconsistency: Consider an inverse problem, where we are unable to fit the data with the chosen parameterization and the available forward relation.
The reason for this may be `noise', defined as the data residual after data fitting.
In this example our problem is overdetermined, which means that the image of the model space (in data space) under the forward function has lower dimensionality than the data space. Consequently, the evidence needed to calculate hyperparameters is an integral over the data distribution, conditioned on a low-dimensional subspace in the data space. This evidence calculation can be made over exactly the same data, and with exactly the same uncertainties, but under different data parameterizations. As represented in Figure~\ref{fig:Conditionals-in-2-frames}, the BK-inconsistency will generally then cause the results to differ under different parameterizations, rendering the calculation meaningless.

We now rigorously investigate the impact of the BK-inconsistency on Hierarchical/empirical Bayesian inversion. We do this through analytical counterexamples and demonstrate that a simple reparameterization of data leads to different results.

Consider a simple inverse problem $\bfd = g(\bfm)$ with 3 data values $\bfd^{obs} = (d_1, d_2,d_3)$, and two model parameters $\bfm = (m_1,m_2$). We assume that the prior is constant in a 2D cube $\calC_{\calM}$ with edge length $\Delta m$, and that the problem is linear in Cartesian coordinates with uniformly distributed uncertainties of $\bfd$, all with the same, unknown uncertainty $\sigma$.

To expose the BK-inconsistency in this problem, we derive the data uncertainty $\sigma$ analytically, by maximizing the evidence for three different parameterizations of data. The data transformations are chosen such that the data and their uncertainties remain the same in all cases. In other words, we solve the same physical problems in three different ways, and arrive at three different solutions for $\sigma$ (see Appendix \ref{Appendix: BK-incon-Hierarch} for details).

\bigskip\noindent
In Case 1 we consider the linear problem
\begin{equation}
    \begin{pmatrix}
    d_1\\
    d_2\\
    d_3
    \end{pmatrix} =
    \begin{Bmatrix}
    0 & a\\
    b & 0\\
    c & 0
    \end{Bmatrix} 
    \begin{pmatrix}
    m_1\\
    m_2
    \end{pmatrix}
    \label{lin-probl-2-Hierarch}
\end{equation}
with $d_1 = 1.5$, $d_2 = 1.1$, $d_3 = 0.9$, $a = 1.0$, $b = 1.0$, and $c = 0.5$, and obtain the solution
\begin{equation}
    \sigma_1 \approx 0.466667 \ 
\end{equation}
by maximising the evidence with respect to $\sigma$ according to hierarchical Bayes.

\bigskip\noindent
In Case 2 we make the trigonometric transformation $d_1 \rightarrow \tan(d_1)$ of the first datum (and its uncertainties), and obtain a different solution:
\begin{equation}
    \sigma_2 \approx 1.02932 \ .
\end{equation}
Finally, we consider the transformation of the first datum to `energy': $d_1 \rightarrow d_1^2$ and obtain the result
\begin{equation}
    \sigma_3 \approx 1.5 \ .
\end{equation}
It is clear that, since we have analysed exactly the same data in all three examples, we have revealed a contradiction, rendering the computed standard deviations, and the evidences used to obtain them in Appendix \ref{Appendix: BK-incon-Hierarch}, meaningless. Such trignometric and energy style transforms (amongst others) are common in physical sciences, illustrating that the implications for hierarchical problems are myriad. 

\subsubsection*{Consequences for hypothesis testing and decision making}
The consequences of the above results for hypothesis testing are serious, as we now demonstrate.

\bigskip\noindent
In Case 1, with optimal data uncertainty $\sigma_1 \approx 0.466667$, both the prior and the likelihood are constant in the square $\calP(\sigma_1)$, so the posterior probability density in parameter space is:
    \begin{align}
       p_{\sigma_1}(\bfm|\bfd_{obs}) = 
    \begin{cases}
    \frac{1}{A(\sigma_1)}
           & {\rm for} \ \bfm \in \calP(\sigma_1)  \\ 
    0      & \text{otherwise}
    \end{cases}
    \label{eq: post-Case-1}
\end{align}
where $ A(\sigma_1) $ is the area of $\calP(\sigma_1) $.

\bigskip\noindent
In Case 2, with optimal data uncertainty $\sigma_2 \approx 1.02932$, the prior is constant, and the likelihood is given by equation (\ref{eq: like-Case-2}). This implies that the posterior probability density in parameter space is:
   \begin{align}
       p_{\sigma_2}(\bfm|\bfd_{obs}) 
    &= \begin{cases}
    \frac{L_2 (\bfm)}{\int_{\calP(\sigma_2)} L_2 (\bfm)\, d\bfm}
           & {\rm for} \ \bfm \in \calP(\sigma_2)  \\ 
    0      & \text{otherwise}
    \end{cases} \\
    &= \begin{cases}
    \frac{a \cos ^2(a m_2)}{\sin (\sigma) \cos (\sigma) \cos (2 d_1)+\sigma}
           & {\rm for} \ \bfm \in \calP(\sigma_2)  \\ 
    0      & \text{otherwise}
    \end{cases} \\
    &\approx \begin{cases}
     1.689 \cos ^2(m_2) & {\rm for} \ \bfm \in \calP(\sigma_2)  \\  
    0      & \text{otherwise}
    \end{cases}
\end{align}
where $ L_2 (\bfm) $ is the likelihood function, and $d_1 = 1.5$, $a = 1.0$.

To illustrate the consequences of the different posteriors on hypothesis testing and decision making, consider a case in which we will make a decision based on whether parameter $m_2$ might exceed some critical threshold of 1.6. We must therefore evaluate the posterior probability that $m_2 > 1.6$. Alternatively, we might test the hypothesis that $m_2 < 1.6$ with 95\% confidence.

\bigskip\noindent
For Case 1 we find that
\begin{equation}
    P^{(1)}_{post}(m_2 > 1.6) = 0
\end{equation}
since the posterior (\ref{eq: post-Case-1}) is zero for $m_2 > (d_1 + \sigma_1)/a \approx 1.566667$.

\bigskip\noindent

\bigskip\noindent
For Case 2, on the other hand, we obtain
\begin{align}
     P^{(2)}_{post}&(m_2 > 1.6) = \int_{1.6}^{(d_1 + \sigma_3)/a} \int_{(d_3-\sigma_3)/c}^{(d_2+\sigma_3)/b} p_{\sigma_2}(\bfm|\bfd_{obs}) \, dm_1 \, dm_2 
     \ \approx \ 0.107  \ .
 \end{align}
In Case 1, the decision can be made safe in the knowledge that $m_1<1.6$ with 100\% certainty, whereas in Case 2 it can not. Under the hypothesis-testing scenario, in Case 1 the hypothesis stands with 100\% certainty, whereas in Case 2 it is only proven to 89.9\% confidence, which is insufficient for most hypothesis tests.

It is clear from these examples that when using Hierarchical Bayes, a simple transformation of data parameters  can lead to different decisions, and to hypotheses being either accepted or rejected, despite the fact that the data and forward functions remain the same.

\subsection*{Trans-Dimensional Inversion and the BK-Inconsistency}
A trans-dimensional inverse problem is one where the dimensionality $k$ of the parameter space is included as a hyperparameter in the set of unknown variables. Hence, $\bfh = k$ in equation \ref{eq: Bayesian-model-select}. 
%
%
The aim of trans-dimensional inversion is 
to find the most favorable dimensionality $k$, measured through its evidence $p_d(\bfd | k)$ (see equation \ref{eq: Bayesian-model-select}), and to compute the corresponding conditional posterior density $p_{m|d,k}(\bfm | \bfd, k)$.

This evidence-based method is responsible for so-called {\em natural parsimony} (the postulate that if a low- and a high-dimensional model have similar data fits, the low-dimensional model is the most probable). In the following we show that in fact, due to the BK-inconsistency, the evidence contains no information about how well a theory or parameterization predicts the data.

\subsubsection*{BK-inconsistent evidences in trans-dimensional inversion}
We now investigate the impact of the BK-inconsistency on trans-dimen\-sional inversion
showing that a simple reparameterization of data leads to two different results.

\bigskip\noindent
Consider a linear inverse problem
\begin{equation}
\bfd = g(\bfm) = \bfG \bfm 
\end{equation}
with 3 data values $\bfd^{obs} = (d_1, d_2,d_3)$, and two model parameterizations, one with $1$ parameter $\bfm = m$ and matrix $\bfG_1$, and the other with $2$ parameters $\bfm = (m_1,m_2)$ and matrix $\bfG_2$. Assume further that the prior is constant in the interval $\calR_1 = [0,{\Delta m}]$ for $1$ parameter, and in the square $\calR_2 = [0,{\Delta m}]\times [0,{\Delta m}]$ for $2$ parameters. 
To expose the BK-inconsistency in this problem, we solved this problem with two different representations of the same data, first in Cartesian, then in spherical coordinates. 
In the concrete example with
\begin{equation}
    \begin{pmatrix}
d_1\\
d_2\\
d_3
\end{pmatrix} =
\begin{pmatrix}
3.1\\
5.8\\
1.1
\end{pmatrix} \ , \  \quad\quad
\bfG_1 = 
\begin{Bmatrix}
2 & 1\\
4 & 2\\
1 & 0
\end{Bmatrix}  \ , \  \quad\quad
\bfG_2 = 
\begin{Bmatrix}
2 \\
4 \\
1 
\end{Bmatrix}
\end{equation}
we computed the evidences for $k = 1$ and $k = 2$ with a Cartesian data parameterization, and obtained (see details in Appendix \ref{Appendix: BK-Inconsist-TransD}): 
\begin{equation}
p(\bfd_{obs}|1) = \frac{1}{(2\sigma)^3} \, \frac{15}{100} \, \frac{1}{\Delta m} \ ,
\end{equation}
and
\begin{equation}
p(\bfd_{obs}|2) =  \frac{d_2 - 2d_1 + 3\sigma}{(2\sigma)^2 {\Delta m^2}}  \ ,
\label{eq: evidence-2D-Cart}
\end{equation}
respectively. For $\Delta m = 2$ and and $\sigma = 0.4$  this gives a Bayes factor of $B = 2.13333\dots$, favoring $k = 2$.
A subsequent solution of the problem with exactly the same data, but now transformed into spherical coordinates, 
resulted in a Bayes factor of $B \approx 0.687\dots$
favoring $k = 1$, in contrast to the Cartesian case.

\subsubsection*{How is this contradiction connected to the BK-inconsistency?}
We have shown that the `optimal' number of model parameters was different in the two, equivalent data parameterizations. To explain this contradiction, let us revisit the definition of the evidence in equation (\ref{eq:evidence}). A closer look reveals that it contains the ``marginal likelihood'' $p_{\bfd}(g_k(\bfm^{(k)}))$ which is the term that is believed to measure the ability of the parameterization to explain the data. However, if the dimensionality of the data space is $N$, and if we consider sparse parameterizations which by definition have dimensionality $M < N$, the image of the model space under the forward function $g_k$ has a dimensionality smaller than $N$. This means that the marginal likelihood $p_{\bfd}(g_k(\bfm^{(k)}))$ is, in fact, a conditional probability density ($p_{\bfd}$ conditioned on $g(\calM))$ in the data space. As a result, this critical term suffers from the BK-inconsistency: The Jacobian transformation, which adjusts the probability density in data space to a new $N$-dimensional data parameterization, does not work in the lower-dimensional image space $g(\calM)$. Hence, the marginal likelihood and its integrals depend critically on the parameterization of the data.
Consequently, the Bayes factor also changes after reparameterization of the data, and thus looses its meaning.

We are led to conclude that trans-dimensional results are incorrect in cases where they are expected to make a difference, namely when the number of model parameters is smaller than the number of data.

\subsection*{Conclusion and Discussion of the impact of the BK-inconsistency on Bayesian Model Selection}
The above result forces us to conclude that standard Bayesian theory does not allow useful comparison between different data- and model priors, and between models of different dimensionalities. 
%
%
Bayesian model selection methods are based on the idea that lack of knowledge about prior data and parameter uncertainties, or incomparability of probabilities between model spaces of different dimensionalities, can be offset by directing our attention to the data fits (evidences). Calculation of evidences involves an integration over the model spaces, thereby apparently removing the differences in dimensionalities, either in the model space, or in its image in the data space. However, as we have seen above, this is not the case. The BK-inconsistency sets in when evidences are calculated in any overdetermined case resulting from either a simplified forward function (as in many applications of hierarchical Bayes), or from the use of sparse parameterizations (as in trans-dimensional inversion). 

In the latter case, the concept of natural parsimony is rendered invalid.
In the literature, there are a number of examples of physical problems which indicate that parsimony is not natural. In a recent work on trans-dimensional electrical resistivity tomography using the reversible-Jump Monte Carlo algorithm  \cite{Galetti2018} it was found that natural parsimony was not achieved for certain choices of step size in the model proposal when changing dimensionality. Furthermore, in a study of trans-dimensional inversion of surface waves \cite{Zhang2018} a clear dependence of the calculated model complexity (number of parameters) on the choice of parameterization of the subsurface was found. These observations comprised important motivations for our investigation, and are both aligned with the findings of this paper.

Our findings also have far-reaching implications for comparisons between different, overlapping, but incompatible model parameterizations. In current Bayesian formulations the only way to compare such parameterizations is through evidence calculations. Our results show that such evidences are not uniquely defined when applied to problems in physical sciences.

\subsection*{Logical Acausality of Hierarchical and Empirical Bayes}
In hierarchical Bayes we have augmented the usual unknown parameters $\bfm$ with  hyper-parameters $\boldsymbol{\theta}$, parameterizing the prior distribution in the data space and in the parameter space. In cases where $\boldsymbol{\theta}$ is unknown, this version of Bayes theorem offers a way to estimate them from the information in data and in the forward function.

This formulation of Bayesian inversion is based on the assumption that there exists a {\em true} (independent of the analyst), but unknown distribution $p_{m|\theta}(\bfm|\boldsymbol{\theta})$ belonging to a parameterized family of distributions which can take the role of prior in the calculation. The parameters $\boldsymbol{\theta}$ defining this prior are themselves considered random with distributions (the so-called {\em hyper priors}) supplied by the analyst(s). The rationale is that, in the classical Bayes method, the prior may be difficult to define by the user, and different users may suggest different priors and obtain different results. The thinking is therefore that there is a randomness in the population of analysts that can be modeled using the hyper-parameters, and if we can provide a prior distribution for the hyper-parameters, we can find their posterior distribution through Bayesian data analysis. In other words, the random population of analysts have an uncertain idea of the `true prior', but this can be improved by the introduction of data.

This basic assumption of hierarchical and empirical Bayes, when applied to inverse problems, contradicts our fundamental Assumption A that the prior is {\em input} to the calculation, originating from the analyst(s). It is true that there is a new (hyper) prior $p_{\theta}(\boldsymbol{\theta})$ to be provided by the analyst(s), but this prior is not about a physical parameter, and is therefore hardly informed by any empirical knowledge -- or even intuition -- as a regular prior on data or model parameters would be. 


The above considerations may seem purely philosophical, but as we show in Appendix \ref{Appendix: Simple-Hierarch-B}, the deviant assumptions in hierarchical and empirical Bayes have concrete mathematical consequences for the analysis. In this appendix we apply a simple invariance argument to demonstrate that the priors computed through hierarchical/empirical Bayes depend on the forward function, and are therefore not priors, but posteriors -- with serious consequences, shown below.

\subsubsection*{Interpreting the results of hierarchical Bayes} 

Even if a way to remove the BK-inconsistency can be found, it is natural at this point to ask questions like: If in hierarchical Bayes we are not computing the prior distribution of the noise, what are we computing? 

\begin{figure}[h!]
\begin{center}
\includegraphics[width=3.1in]{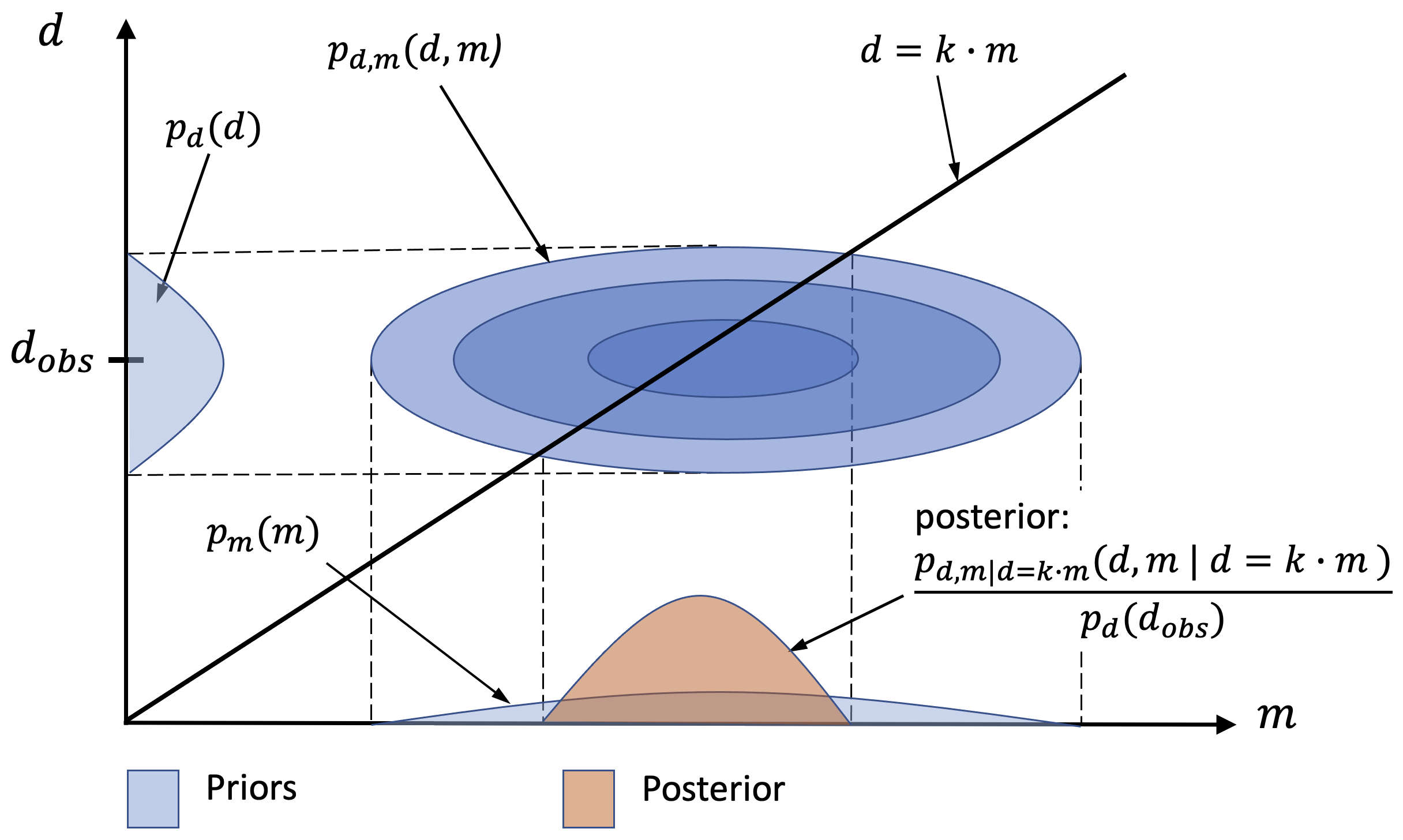}
\quad
\includegraphics[width=3.1in]{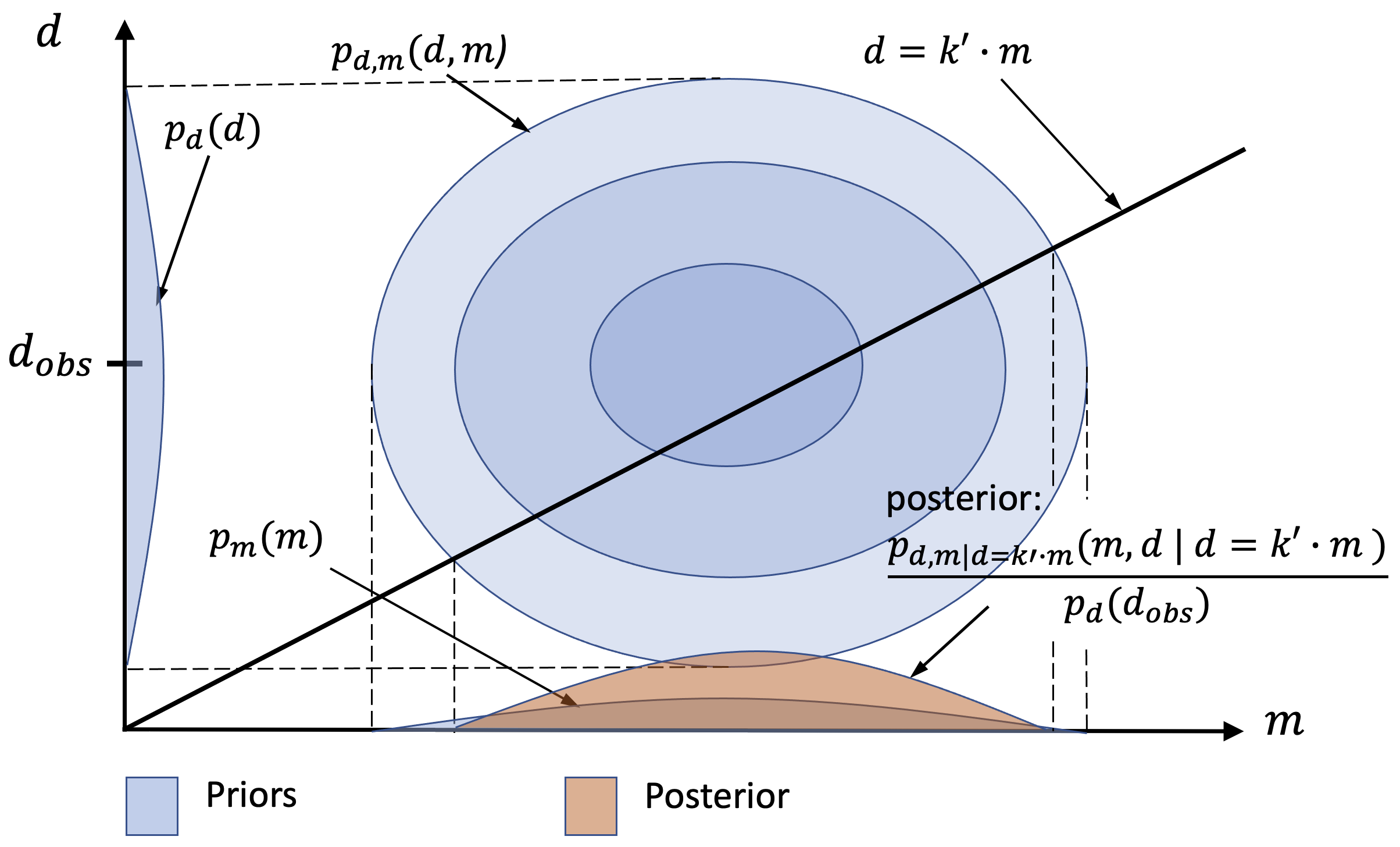}
\caption{\small Noise distributions generated by Hierarchical Bayes for two different forward relations. The joint prior probability density $p_{d,m}$ is the product of the prior probability densities $p_m$ and $p_d$, and $g_m(\bfm)$ is a linear forward function with different gradients in left and right panels. In each panel, the prior on $\bfm$ is fixed, whereas the prior data variance is described by a hyperparameter. In each case, the integral over the joint prior $p_{m,d}(m,d)$, along the subspace defined by the forward function, is maximized with respect to the data variance, resulting in different data variances and different posteriors for the different forward functions.}
\label{fig:data-uncert-forw-depend}
\end{center}
\end{figure}

\begin{figure}[h!]
\begin{center}
\includegraphics[width=3.5in]{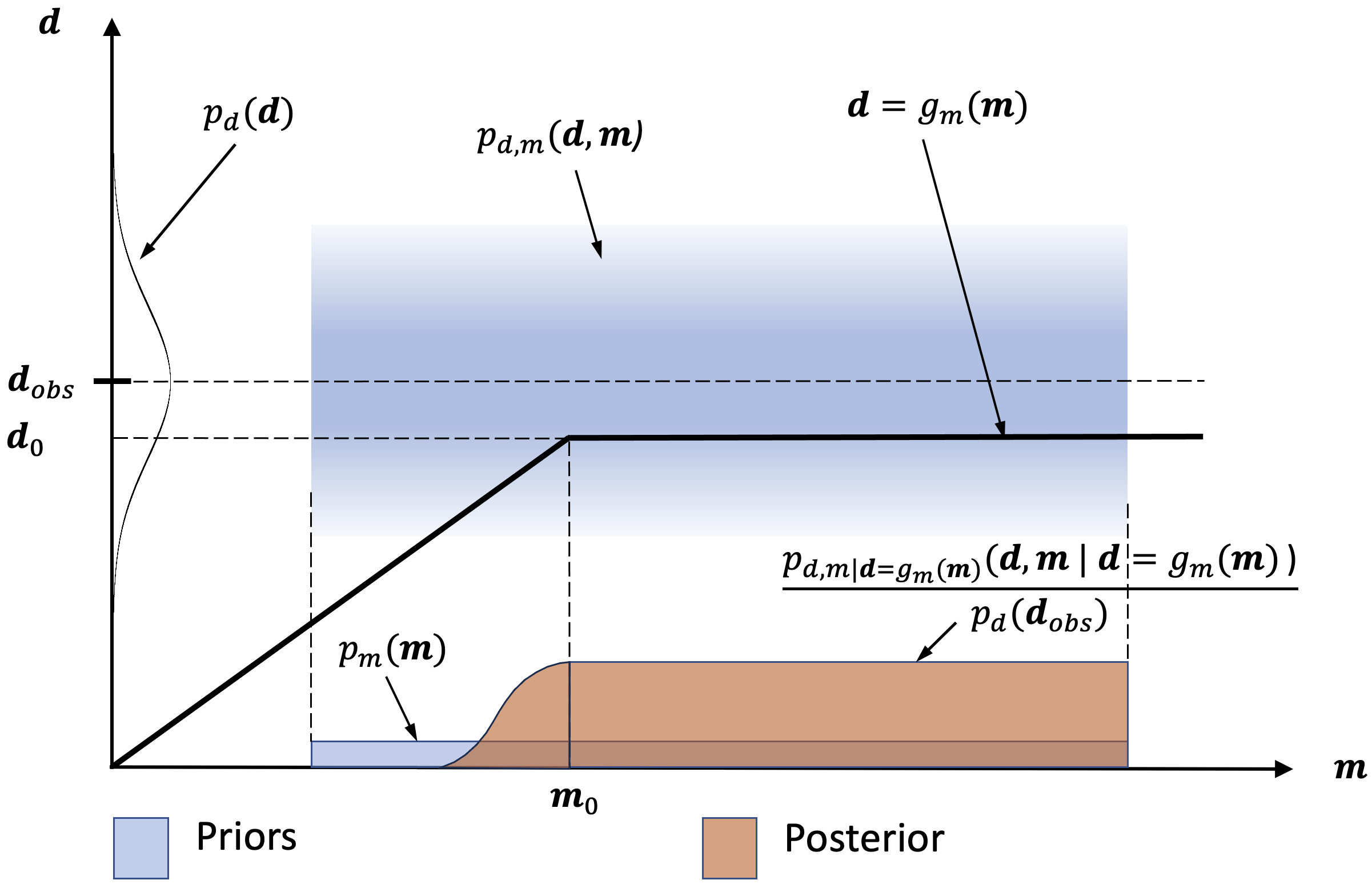}
\caption{\small Hierarchical inference in a case where the forward function is unable to fit the data. The joint prior probability density $p_{d,m}$ is the product of the prior probability densities $p_m$ and $p_d$, and $g_m(\bfm)$ is the forward function. In this case, hierarchical Bayes will produce a standard deviation for the data that maximizes the integrated posterior probability over the model space.}
\label{fig:Unmodeled-data}
\end{center}
\end{figure}
This can be understood using figure \ref{fig:data-uncert-forw-depend}: hierarchical Bayes favors values of hyperparameter ${\boldsymbol\theta}$ that maximize the integral over the joint prior $p_{m,d}(\bfm,\bfd)$, along the subspace defined by the forward function. When the forward function changes, a maximum occurs at different values of ${\boldsymbol\theta}$, resulting in different data variances and different parameter space posteriors.

To see the effect of this when the model parameters are unable to fit the observed data, Figure \ref{fig:Unmodeled-data} shows data with Gaussian uncertainties with the observations as mean, and an unknown standard deviation. In this case, hierarchical/empirical Bayes will find a data variance maximizing the integral over the joint prior $p_{m,d}(\bfm,\bfd)$ for $\bfd=g(\bfm)$. This maximum obviously exists, and will depend strongly on the forward function (and the parameter prior). The optimal variance must increase as $\bfd_{obs}$ is further from (above) the forward function plateau. So, given $\bfd_{obs}$, the variance measures how poorly the forward function is able to fit the data, and the result is entirely {\em independent} of  prior uncertainty on the data (the observational noise). Our conclusion is that in this case ${\boldsymbol\theta}$ is a measure of the {\em modelization error}, rather than an estimate of observational noise. 

Perhaps it sounds useful that the estimated `noise' is the part of the data that originates from structure that cannot be represented by our parameterization, and/or our forward function. However, in cases of real inconsistency between observed data, and data computed from a priori reasonable models, the above examples demonstrate that this fact may remain hidden by hierarchical/empirical Bayes. Instead, inconsistencies will tend to be absorbed by an increase in the estimated data uncertainty, making it more difficult for given data to reject a model, or equivalently, a hypothesis. This is, from a scientific perspective, a serious problem.

\section*{Closing Remarks}

The results above demonstrate inconsistency in over-determined problems. This arises because the image of parameter space through the forward function lies in a strict subspace of the data-space, illustrated schematically in Figure~\ref{fig:Borel-in-2-frames}. However, Figure~\ref{fig:Borel-in-2-frames} deliberately has no axis labels: since the BK-inconsistency applies to any subspace which is conditioned on values of its complement, the axes represent any of (i) two different data, (ii) two different parameters, or (iii) one of each. While interpretation (i) has been discussed in detail above, consider a problem in which only a subset of parameters are of interest, the others being considered nuisance parameters. Interpretation (ii) of Figure~\ref{fig:Borel-in-2-frames} then illustrates that the probabilities of any parameter of interest (in red), depend on the {\it parametrization} of any nuisance parameter (vertical axis), a result that is not widely acknowledged. A related problem occurs when using Bayesian dimensionality reduction methods. For example, in the field of machine learning, generative adversarial networks \cite{Goodfellow2014}, variational autoencoders \cite{Kingma2014} and a variety of other algorithms are commonly applied to find a (usually nonlinear) lower dimensional subspace of parameter space that contains important components of parameter variability. When results are interpreted probabilistically, they suffer from the same inconsistency: interpretation (ii) applies to subspaces of any geometry, so probability distributions on that subspace depend on the parametrization of the rest of parameter space. Finally, consider a 1-parameter and 1-datum system in which Bayes theorem constructs the conditional distribution $p_{m|d} (\bfm | \bfd)$: under interpretation (iii) with the datum on the vertical axis, Figure~\ref{fig:Borel-in-2-frames} illustrates that this conditional is inadmissible -- as is $p_{d|m} (\bfd | \bfm)$ if we exchange the axis labels. Bayes theorem itself is thus inconsistent, a problem that is thus ingrained within {\it all} standard Bayesian theory or practice that invokes equation~\ref{eq:Bayes-precise}.

We have shown that when parameterizations are sparse, the marginal likelihood, or evidence, suffers from the BK-inconsistency so that results vary with changes in model parameterizations. This has implications for frequently applied statistical tests and criteria. For example, the Akaike Information Criterion \cite{akaike1974new} states that a suitable measure to assess the quality of one hypothesis or model (defined by hyperparameters $\theta$) over another is 
\begin{equation}
AIC(\theta) = 2k - 2\ln{(\text{maximum likelihood})}   
\label{eq:AIC}\ 
\end{equation}
where $k$ defines the number of parameters in each model $\bfm$ given $\theta$. The preferred model is that which minimises $AIC(\theta)$. However, expression \ref{eq:AIC} is derived by comparing the parametrization defined by $\theta$ to that of the `true', or perfectly parametrized model \cite{akaike1974new}, assuming that the former has fewer parameters than (defines a subspace of) the latter. By definition, the $AIC$ is then subject to the BK-inconsistency when there are more data than parameters. This deficiency is also visibly manifest in expression~\ref{eq:AIC}: it includes the maximum likelihood estimator of model $\bfm$, which we have shown suffers from the BK-inconsistency because it resides in a zero-dimensional sub-space of model parameter space (Appendix~\ref{Appendix: Inconsistent-MAP}).

We demonstrated that whether hypotheses pass or fail standard tests of statistical significance can be changed simply by altering the parametrization of the data. Parametrizations are rarely unique, so this requires that we re-examine significance tests in physical problems. For example, in clinical testing of novel medical treatments \cite{Goligher2024}, a simple change from the variables describing effects to the reciprocals of those variables might (be used to) change statistical outcomes, similarly to the analysis in Appendix~\ref{Appendix: Tomo-Ex}, even though outcomes in the real world are unaltered.  Hierarchical methods are also used in clinical trials \cite{McGlothlin2018Bayesian,Goligher2023}, yet we have shown that hierarchical and empirical Bayes can produce results that depend primarily on the setup of the problem rather than on information in recorded data (appendices~\ref{Appendix: Simple-Hierarch-B} and~\ref{Appendix: BK-incon-Hierarch}). This produces perverse incentives when a failed hypothesis test would block progress.

One possible solution might seem to be to fix the parametrization of a problem, and calibrate the outcomes of decision-making to the implied metric space. For example, previous experience may have established that satisfactory performance of a treatment is observed if it has passed established tests using a standardized parametrization. If future treatments produce similar test results using the same parametrizations, can we be assured of similar treatment performance? The answer is, no: consider a treatment who's outcome depends both on the patient's physical state and their socioeconomic situation (which, when both are parametrized, define the data space) \cite{Iqbal2020}. We have shown that the significance of physical outcomes (a data subspace) may change with the parametrization of their social demographic (a different subspace), as illustrated in figure~\ref{fig:Borel-in-2-frames}. As new factors with new parametrizations emerge with future research, this putative solution fails.

Similar considerations apply to Bayesian decision-making problems \cite{Fenton2019} in which changes of variables that describe potentially hazardous effects could lead to decisions (models, hypotheses or treatments being accepted or rejected) such that a desired utility decreases rather than increases. And generally any method for scientific conjecture and testing which invokes Bayes theorem may lead to incorrect interpretations of outcomes.

While this article mainly addresses Bayesian statistical analysis, a frequentist approach may also be expected to lead to different outcomes under a change of parametrization. Frequentist analysis relies entirely on the probability of observing data under a null hypothesis, which may be written in the form $P$(data$|$hypothesis), although under a different interpretation compared to the likelihood functions above \cite{Goligher2024}. Such conditional distributions are also subject to the BK-inconsistency, so a similarly critical analysis of frequentist methods is warranted.

More generally, it is already known that either a Bayesian or frequentist analysis of data might not reject a null hypothesis concerning a treatment or action for two reasons: the action has no or minimal effect, or because the information in the data was insufficient. To this we now add a third possibility: that the parametrization of the system (either parameters or data) conferred a bias towards the null hypothesis.

A formulation of probabilistic inversion, proposed to avoid the BK-inconsistency, was presented in \cite{Tarantola82a}, and later in \cite{Mosegaard-Tarantola-2002} a way of avoiding the inconsistency by explicit introduction of the metric in data and model spaces was proposed. These methods are now rarely invoked, and it is not at all clear how they could remedy the inconsistencies in trans-dimensional model selection as described here.

This study therefore leads us to conclude that careful rethinking of Bayesian inversion practices is required, and a similar analysis of frequentist methods should be undertaken. We have seen that the promises of Bayesian model selection methods cannot be kept, namely that information previously considered independent and indispensable (model structure, priors in data and model space) can be computed from the data. It cannot be ruled out that the inconsistencies we have exposed in this paper have a partial solution, but what that solution is, is currently unclear.

\section*{Acknowledgements}
Klaus Mosegaard would like to thank Domenico Giardini, Amir Khan, Andrew Jackson and their colleagues at the Department of Earth and Planetary Sciences, ETH Z{\"u}rich, for a stimulating working environment during his stay in the spring 2024. Thanks also to the Inge Lehmann Grant committee under The Royal Danish Academy of Sciences and Letters for supporting Klaus Mosegaard during his visit at ETH Z{\"u}rich.

\bigskip

\bibliographystyle{sciencemag}
\bibliography{BayesModSelectReferences}{}

\clearpage



\clearpage


\begin{appendices}


\section{A simple tomographic example}
\label{Appendix: Tomo-Ex}
We consider a tomographic example where data $\bfd$ and unknown parameters $\bfm$ are connected through the forward relation
\[
\bfd = g(\bfm) \ .
\]
Here, $\bfd$, as well as $\bfm$, has 2 components.
$\bfd$ contains the arrival times along two straight-line rays passing through a medium with 2 homogeneous blocks of width $l=1$ unit, characterized by their wave speeds $v$ or their wave slownesses $s$ (see figure \ref{fig:Simple-tomo}).
Using velocity parameters, the uncertainty of $\bfd$ has a uniform probability density $p_d$ on a 2-D interval of the data space $\calD$, and the prior on $\bfm$ is uniform on a 2-D interval in the model space $\calM_v$ (see figure \ref{fig:Bayes-uniform}).  The joint prior over model and data space is therefore uniform and non-zero within the corresponding bounds.

\subsubsection*{Case 1: Velocity inference}
In this case the unknowns are velocity parameters $\bfv$, with a uniform prior $p_m (\bfv)$ over a subset $\calM_v$ of the model space. In other words, if the forward relation is $\bfd = g_v (\bfv)$, the problem has the simple structure shown in figure (\ref{fig:Bayes-uniform}). For this problem, using Formulation 2 of Bayes Theorem (equations \ref{eq: Bayes-2} and \ref{eq: Bayes-2alt}), the posterior probability density becomes: 
\begin{equation}
    p_{v|d} (\bfv | \bfd_{obs}) = \frac{p_{d,v}(\bfd,\bfv|\calG)}{p_d(\bfd_{obs})} = 
\begin{cases}
    K' & {\rm for} \ [g_v(\bfv) \in \calD] \wedge [\bfv \in \calM_v]  \\ 
    0              & \text{otherwise}
\end{cases}
\label{eq:Posterior-v-A}
\end{equation}
where $K'$ is a normalization constant. Thus, the posterior $p_{v|d} (\bfv | \bfd_{obs})$ is homogeneous, since it is equal to the homogeneous, joint prior on $\calM_v \times \calD$, conditioned (evaluated) on the manifold given by $\bfd=g_v (\bfv)$, and normalized so that it integrates to $1$. We emphasize that in Formulation 2, $\bfd$ is a random variable and contains the measurement noise.

Say we now are interested in the inversion result under the additional constraint that the velocities are the same in the two blocks. From (\ref{eq:Posterior-v-A}) we compute the {\em conditional} posterior on the 1-D subspace defined by $v_2 = v_1$, as a function of $v_1$:
\begin{equation}
\begin{aligned}
    p_{v_1|d} (v_1 | \bfd_{obs}) &= p_{v|d, v_2 = v_1} (\bfv | \bfd_{obs}, v_2 = v_1) \\
    &=   
\begin{cases}
    C' & {\rm for} \ [g_v({v_1,v_1}) \in \calD] \wedge [({v_1,v_1}) \in \calM_v]  \\ 
    0              & \text{otherwise}
\end{cases}
\end{aligned}
\label{eq:Cond-Posterior-v-A}
\end{equation}
where $C'$ is  constant. This density is, of course, constant over an interval.

\subsubsection*{Case 2: Slowness inference}
Let us now reparameterize from velocities to slownesses $\bfv = (v_1,v_2)^T \rightarrow \bfs = (s_1,s_2)^T$ through the transformation $h$:
\begin{align}
         \begin{pmatrix}
           s_1 \\
           s_2 
         \end{pmatrix}
       = h\begin{pmatrix}
           v_1 \\
           v_2 
         \end{pmatrix}
       = \begin{pmatrix}
           1/v_1 \\
           1/v_2 
         \end{pmatrix}
\label{eq:v-to-s-A}
\end{align}
The Jacobian matrix of the transformation is:
\begin{align}
         \bfJ =          
         \begin{Bmatrix} \frac{\partial s_i}{\partial v_j}\end{Bmatrix} =
         \begin{Bmatrix}
           -1/v_1^2 & 0 \\
           0 & -1/v_2^2 \\ 
          \end{Bmatrix}
\label{eq:Jacobian-s-to-v}
\end{align}
whose absolute value of the determinant is
\begin{equation}
|\det(\bfJ)| = \frac{1}{v_1^2 v_2^2} \ = s_1^2 s_2^2
\label{eq:det-Jacobian-s-to-v}
\end{equation}
Hence, the transformed prior over slowness becomes:
\begin{equation}
   p_s(\bfs) = |\det(\bfJ)|^{-1} p_m(\bfv(\bfs))  \propto 
   \begin{cases}
           s_1^{-2} s_2^{-2} & {\rm for} \ (s_1,s_2) \in \calM_s \\
           0 & \text{otherwise}          
   \end{cases}
\end{equation}
where $\calM_s = h(\calM_v)$, and the joint prior distribution over $\bfd$ and $\bfs$ takes the same values as $p_s(\bfs)$, distributed uniformly across $\calD$ and renormalised.  If, in the new parameterization, our forward relation is
\[
\bfd = g_s(\bfs)  \ 
\]
then the transformed posterior probability density for $\bfs$ is the joint prior on $\calM_s \times \calD$, conditioned on the manifold given by $\bfd = g_s(\bfs)$, and normalized by $p_d(\bfd_{obs})$, is (using again Formulation 2 of Bayes Theorem):
\begin{align}
    p_{s|d_{obs}} (\bfs | \bfd_{obs}) &= \frac{p_{d,s}(\bfd,\bfs|\bfd = g_s(\bfs))}{p_d(\bfd_{obs})} \\
    &\propto 
\begin{cases}
           s_1^{-2} s_2^{-2} & {\rm for} \ [g_s(s_1,s_2) \in \calD] \wedge [(s_1,s_2) \in \calM_s] \\
           0 & \text{otherwise}          
\end{cases}
\label{eq:Posterior-s-A}
\end{align}
From (\ref{eq:Posterior-s-A}) we can now -- again -- compute the posterior, conditioned on  $v_2=v_1$, or  equivalently: $s_1=s_2$, this time written as a function of $\bfs$. We obtain the 1-D density
\begin{equation}
    p_{s_1 |d} (s_1 | \bfd_{obs}) \propto  
\begin{cases}
           s_1^{-4} & {\rm for} \ g_s(s_1,s_1) \in \calD \wedge (s_1,s_1) \in \calM_s \\
           0 & \text{otherwise}          
\end{cases}
\label{eq:Posterior-s1}
\end{equation}
Transforming this back to the velocity space gives (when formulated as a function of $v_1$): 
\begin{equation}
\begin{split}
p_{v_1|d} (v_1 | & \bfd_{obs}, v_2 = v_1) = \left|\frac{dv_1}{ds_1}\right|^{-1} p_{s_1|d}(s_1|\bfd_{obs}) \\ \propto 
 &\quad  
 \begin{cases}
    v_1^2 & {\rm for} \ g_v({v_1,v_1}) \in \calD \wedge ({v_1,v_1}) \in \calM_v  \\ 
    0  & \text{otherwise}
\end{cases}
 \\
 \end{split}
\label{eq:Cond-Posterior-v-alt-A}
\end{equation}
which is in direct contradiction to expression (\ref{eq:Cond-Posterior-v-A}).

\clearpage

\section{Inconsistency of MAP solutions}
\label{Appendix: Inconsistent-MAP}
Assume that we wish to find a maximum likelihood solution, or a maximum posterior solution, to an inverse problem. As we shall see, this runs into the BR-inconsistency, because the maximum posterior value is the posterior probability density conditioned to a point (a zero-dimensional subspace).

\bigskip\noindent
Imagine that we have analyzed laboratory data, and solved an inverse problem leading to a solution for temperature $T$ and density $\rho$ of a gas in a confined volume. We have found that the joint posterior probability density for the two parameters is a product of two log-normal distributions:

{\small
\begin{equation}
f(T,\rho) = 
\frac{1}{T\sigma_T \sqrt{2\pi}}\exp\left( -\frac{\left(\ln(T)-\mu_T\right)^2}{2\sigma_T^2} \right)
\frac{1}{\rho\sigma_\rho\sqrt{2\pi}}\exp\left( -\frac{\left(\ln(\rho)-\mu_\rho\right)^2}{2\sigma_\rho^2} \right).
\end{equation}
}It can be shown that the maximum point for this distribution (the ``best-fitting solution'' to the inverse problem) is:
\begin{equation}
(T,\rho)_{max} = 
\left(\exp\left(\mu_T-\sigma_T^2\right),\exp\left(\mu_\rho-\sigma_\rho^2\right)\right).
\end{equation}
Assuming that the gas is an ideal gas, and that we have a fixed amount of gas during the entire experiment, we can express the pressure $p$ as:
\begin{equation}
p = a T \rho
\end{equation}
where $a$ is a constant.

\begin{figure}
\begin{center}
 \includegraphics[width=3.0in]{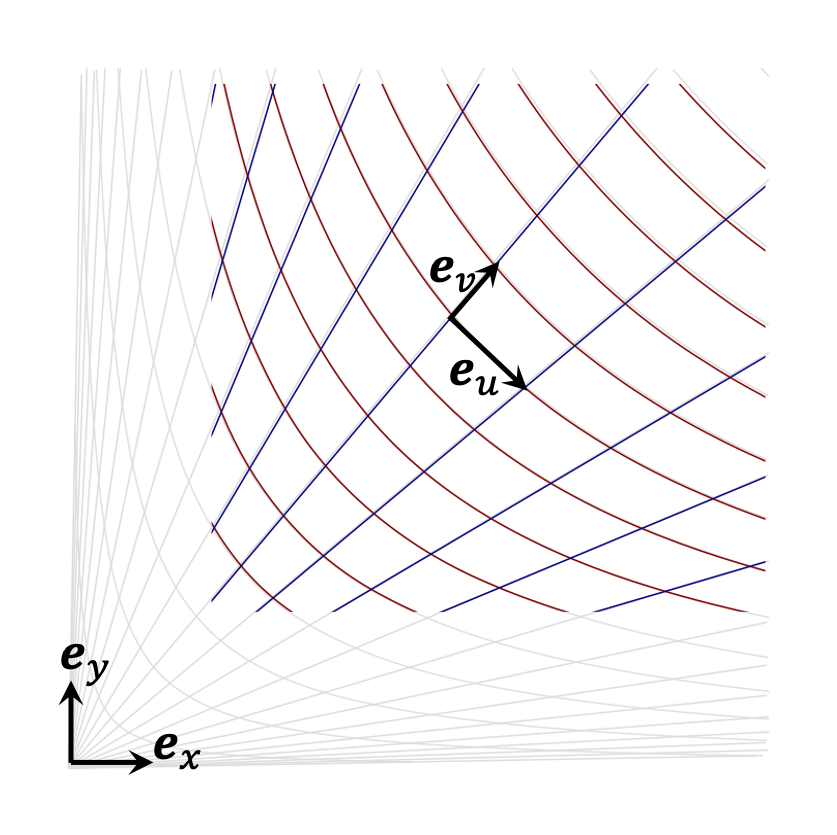}
 \caption{\small Hyperbolic coordinates in the 1st quadrant of the 2D Euclidian space}
 \label{fig: Hada-T}
\end{center}
\end{figure}
To facilitate the subsequent analysis, we choose to change coordinates such that one of the coordinates represents the pressure. The required change into {\em hyperbolic} coordinates $(u,v)$ is as follows:
\begin{equation}
u = \frac{1}{2}\ln\left( \frac{T}{\rho} \right)
\end{equation}
\begin{equation}
v = (T\rho)^{1/2}
\end{equation}
where $v^2$ is proportional to the pressure.
The $x$-$y$ and the $u$-$v$ coordinate systems are equivalent, in the sense that the transformation $h$ is continuous and one-to-one (as it must be).

Using a Jacobian transformation, we can now calculate the posterior probability density $g(u,v)$ in the u-v coordinate system, expressed in $u$-$v$ coordinates:
\begin{equation}
g(u,v) = 
\frac{1}{\pi v \sigma_T \sigma_{\rho}}\exp\left( -\frac{\left(\ln v+u-\mu_T\right)^2}{2\sigma_T^2} 
-\frac{\left(\ln v-u-\mu_\rho\right)^2}{2\sigma_\rho^2} \right).
\end{equation}
It can be shown that the maximum of $g(u,v)$ is achieved at the point
{\scriptsize
\begin{align*}
(u,v)_{max} = \Bigg( \frac{1}{2}\left(\left(\mu_T-\frac{\sigma_T^2}{2}\right)-\left(\mu_T-\frac{\sigma_{\rho}^2}{2}\right)  \right) , 
\exp \left( \frac{1}{2}\left(\left(\mu_T-\frac{\sigma_T^2}{2}\right)+\left(\mu_T-\frac{\sigma_{\rho}^2}{2}\right)  \right) \right) \Bigg)
\end{align*}
}
If we transform this result back to the $x$-$y$ coordinate system, we get the point
\begin{equation}
(\hat{T},\hat{\rho})_{max} = 
\left(\exp\left(\mu_T-\frac{\sigma_T^2}{2}\right),\exp\left(\mu_\rho-\frac{\sigma_{\rho}^2}{2}\right)\right)
\end{equation}
which is not the maximum for $f(T,\rho)$. This counter-example demonstrates that the concept of MAP solutions is inconsistent under a change of parameterisation.

In addition to this, it is easily verified that for $\mu_T = \mu_\rho = \sigma_T = \sigma_\rho = 1$ the maximum likelihood value in the $(T,\rho)$ parameterization is $0.0585$, but in the $(u,v)$ parameterization it is $0.0456$.

\clearpage

\section{BK-inconsistent evidences in hierarchical \\ Bayesian inversion}
\label{Appendix: BK-incon-Hierarch}
\subsection{Case 1: Cartesian data and constant priors}
In this case, the data uncertainty distribution is non-zero and constant in a cube $\calC_{\calD}$ in data space, centered at $\bfd^{obs}$, with edge length $2\sigma$. Outside $\calC_{\calD}$ it is zero. The prior in the parameter space is constant in the 2D interval $\calC_{\calM}$ with edge length $\Delta m$, and zero outside. Furthermore, let us for simplicity assume that no model that fits the data has zero prior probability: $g^{-1}(\calC_{\calD}) \subseteq \calC_{\calM}$.

Consider a simple example with the linear forward function:
\begin{equation}
\begin{pmatrix}
d_1\\
d_2\\
d_3
\end{pmatrix} =
\begin{Bmatrix}
0 & a\\
b & 0\\
c & 0
\end{Bmatrix} 
\begin{pmatrix}
m_1\\
m_2
\end{pmatrix}
\label{lin-probl-2}
\end{equation}
\noindent
where $(d_1,d_2,d_3)^T$ are the observed data with uniformly distributed noise with dispersion $\pm \sigma$. Assuming that $\sigma \geq (bd_3 - cd_2)/(c+b))$, the models that fit all 3 data within their error bars are located in a parallelogram in parameter space with corner points: 
\begin{align}
P_1 &= \left(\frac{d_3 - \sigma}{c}, \frac{d_1 - \sigma}{a} \right) \\
P_2 &= \left(\frac{d_2 + \sigma}{b}, \frac{d_1 - \sigma}{a} \right) \\
P_3 &= \left(\frac{d_2 + \sigma}{b}, \frac{d_1 + \sigma}{a} \right) \\
P_4 &= \left(\frac{d_3 - \sigma}{c}, \frac{d_1 + \sigma}{a} \right) .
\end{align}
The area of this parallelogram is 
\begin{equation}
A(\sigma)  = \left( \frac{d_2 + \sigma}{b}-\frac{d_3 - \sigma}{c} \right) 
    \left( \frac{d_1 + \sigma}{a}-\frac{d_1 - \sigma}{a}\right) \ .
\end{equation}
If $\bfd = (d_1, d_2, d_3)$ and $\bfm = (m_1, m_2)$, then for any particular $\sigma$ the evidence is 
\begin{equation}
p(\bfd_{obs}|\sigma) = \int_\calM L_1(\bfm) p_m(\bfm) d\bfm
\end{equation}
where $L_1 (\bfm) = p_d(\bfG \bfm)$ is the likelihood function (see equation \ref{eq: 2nd likelihood}),
\begin{equation}
   p_d(\bfd) = 
   \begin{cases}
       \frac{1}{(2\sigma)^3} & {\rm for} \ \bfd 
             \in \calC_{\calD} \\
       0 & \text{otherwise}          
   \end{cases} 
\label{eq: data-distrib-cart}
\end{equation}
is the prior in the data space, $p_m(\bfm)$ is the model prior, and in formulation 2, $\bfd$ is a random variable which contains the measurement noise.  Given that the prior is nonzero everywhere inside the rectangular area $\calP(\sigma) = P_1 P_2 P_3 P_4$, it contributes a factor $1/\Delta m^2$ to the evidence, giving
\begin{equation}
p(\bfd_{obs}|\sigma) = \frac{1}{\Delta m^2} \, \frac{1}{(2\sigma)^3} \, A(\sigma)  = 
                  \frac{1}{\Delta m^2}
                  \frac{2 \sigma (b (\sigma-d_3)+c (d_2+\sigma))}{8 \sigma^3 a b c} 
                  \ .
\end{equation}
This evidence is maximized for
\begin{equation}
    \sigma = \frac{2 (b d_3-c d_2)}{b+c}
\end{equation}
which is the data uncertainty favored by the hierarchical Bayes method. If, for example, $d_1 = 1.5$, $d_2 = 1.1$, $d_3 = 0.9$, $a = 1.0$, $b = 1.0$, and $c = 0.5$, we obtain
\begin{equation}
    \sigma \approx 0.466667 \ .
\end{equation}

\subsection{Case 2: Transformed data with $d_1 \rightarrow \tan(d_1)$}
We now reconsider the linear inverse problem in equation \ref{lin-probl-2}, with data and uncertainties transformed as

\begin{equation}
\begin{pmatrix}
y_1\\
y_2\\
y_3
\end{pmatrix} =
T_1\begin{pmatrix}
d_1\\
d_2\\
d_3
\end{pmatrix} =
\begin{pmatrix}
\tan(d_1)\\
d_2\\
d_3
\end{pmatrix} \ ,
\end{equation}
with the inverse
\begin{equation}
\begin{pmatrix}
d_1\\
d_2\\
d_3
\end{pmatrix} =
T_1^{-1}\begin{pmatrix}
y_1\\
y_2\\
y_3
\end{pmatrix} =\begin{pmatrix}
\arctan(y_1)\\
y_2\\
y_3
\end{pmatrix} \ .
\end{equation}
The Jacobian of $T_1^{-1}$ is
\begin{equation}
|\det(\bfJ_S)| = \left| \frac{\partial(d_1, d_2, d_3)}{\partial(y_1, y_2, y_3)} \right| = \frac{1}{y_1^{2}+1} \ ,
\label{Jac-arctan}
\end{equation}
giving the transformed (prior) data distribution:
\begin{equation}
   p_y(\bfy) = 
   \begin{cases}
       \frac{1}{8\sigma^3}\frac{1}{y_1^{2}+1} & {\rm for} \ \bfy \in T(\calC_{\calD}) \\
       0 & \text{otherwise}          
   \end{cases}
   \label{eq: data-prior-arctan}
\end{equation}
where $\bfy = (y_1, y_2, y_3)$, and where we have carried out a Jacobian transformation using equation (\ref{Jac-arctan}). 
The transformed forward function 
\begin{equation}
    \bfy = g_y(\bfm) = 
    \begin{pmatrix}
        \tan(a m_2)\\
        bm_1\\
        cm_1
   \end{pmatrix}
\end{equation}
is now inserted into the data prior (\ref{eq: data-prior-arctan}) giving the likelihood (see equation \ref{eq: 2nd likelihood}):
\begin{align}
 L_2 (\bfm) = p_y(g_y(\bfm))
 &= \begin{cases}
    \frac{1}{8\sigma^3} \cos^2(a m_2) &{\rm for} \ \bfm \in \calP(\sigma)  \\ 
    0      & \text{otherwise\ .}
    \end{cases} 
    \label{eq: like-Case-2}
\end{align}
This likelihood was derived through Formulation 2. For comparison, see Appendix \ref{Appendix: Likelihood-Formulation-1} for a standard derivation (through Formulation 1) of this likelihood.
To obtain the evidence, we integrate the product of $L_2 (\bfm)$ and the prior $1/{\Delta m^2}$ over $\calP(\sigma)$:
\begin{align}
&p_(\bfd_{obs}|\sigma) = \frac{1}{\Delta m^2} \int_{\calP(\sigma)} L_2 (\bfm)\, d\bfm \\
   &= \frac{1}{\Delta m^2} \int_{(d_1-\sigma)/a}^{(d_1+\sigma)/a} 
     \left( \int_{(d_3-\sigma)/c}^{(d_2+\sigma)/b} \frac{1}{8\sigma^3} \cos^2(a m_2)\, dm_1 \right) dm_2 \\
   &= \frac{1}{\Delta m^2} \frac{1}{8\sigma^3}
   \left(\frac{d_2+\sigma}{b}-\frac{d_3-\sigma}{c}\right)
   \left(\frac{\cos (2 d_1) \sin (\sigma) \cos (\sigma)+\sigma}{a}\right)
\label{eq: Sinh-evid-2}
\end{align}
When $d_1 = 1.5$, $d_2 = 1.1$, $d_3 = 0.9$, $a = 1.0$, $b = 1.0$, and $c = 0.5$, the preferred value of $\sigma$, maximizing the evidence, becomes
\begin{equation}
    \sigma_2 \approx 1.02932 \ 
\end{equation}
in contrast to the result in Case 1.

\subsection{Case 3: Transformed data with $d_1 \rightarrow d_1^2$}
We now reconsider the problem in equation \ref{lin-probl-2}, with data assumed positive, and the first datum transformed into `energy':
\begin{equation}
T_1\begin{pmatrix}
d_1\\
d_2\\
d_3
\end{pmatrix} =
\begin{pmatrix}
d_1^2\\
d_2\\
d_3
\end{pmatrix} \ ,
\end{equation}
with the inverse
\begin{equation}
T_1^{-1}\begin{pmatrix}
y_1\\
y_2\\
y_3
\end{pmatrix} =
\begin{pmatrix}
y_1^\frac{1}{2}\\
y_2\\
y_3
\end{pmatrix} \ .
\end{equation}
The Jacobian of $T_1^{-1}$ is
\begin{equation}
|\det(\bfJ_S)| = \left| \frac{\partial(d_1, d_2, d_3)}{\partial(y_1, y_2, y_3)} \right| = \frac{1}{2}y_1^{-1/2} \ ,
\label{Jac-Sinh-Sph}
\end{equation}
giving the transformed (prior) data distribution:
\begin{equation}
   p_y(\bfy) = 
   \begin{cases}
       \frac{1}{16\sigma^3} y_1^{-1/2} & {\rm for} \ \bfy \in T(\calC_{\calD}) \\
       0 & \text{otherwise}          
   \end{cases}
   \label{eq: data-prior-energy}
\end{equation}
where we have carried out a Jacobian transformation using equation (\ref{Jac-Sinh-Sph}). 
The transformed forward function
\begin{equation}
    \bfy = g_y(\bfm) = 
    \begin{pmatrix}
        (am_2)^2 \\
        bm_1\\
        cm_1
   \end{pmatrix}
\end{equation}
can now be used to compute the likelihood (see equation \ref{eq: 2nd likelihood})
\begin{align}
 L_3 (\bfm) = p_y(g_y(\bfm)) = 
    \begin{cases}
    \frac{1}{16 \sigma^3 a} \frac{1}{m_2} 
           & {\rm for} \ \bfm \in \calP(\sigma)  \\ 
    0      & \text{otherwise}
    \end{cases}
    \label{eq: like-case-2}
\end{align}
To obtain the evidence, we now integrate the product of $L_2 (\bfm)$ and the prior $1/{\Delta m^2}$ over $\calP(\sigma)$:
\begin{align}
p_(\bfd&_{obs}|\sigma) = \frac{1}{\Delta m^2} \int_{\calP(\sigma)} L_3 (\bfm)\, d\bfm \\
   &= \frac{1}{\Delta m^2} \int_{(d_1-\sigma)/a}^{(d_1+\sigma)/a} 
     \left( \int_{(d_3-\sigma)/c}^{(d_2+\sigma)/b} \frac{1}{16 \sigma^3 a} \frac{1}{m_2} \ dm_1 \right) dm_2 \\
   &= \frac{\left(\frac{d_2+\sigma}{b}-\frac{d_3-\sigma}{c}\right) \left(\ln \left(\frac{d_1+\sigma}{a}\right)-\ln \left(\frac{d_1-\sigma}{a}\right)\right)}{16 a \sigma^3 \Delta m^2}
\end{align}
When $d_1 = 1.5$, $d_2 = 1.1$, $d_3 = 0.9$, $a = 1.0$, $b = 1.0$, and $c = 0.5$, the value of $\sigma$, maximizing the evidence, becomes
\begin{equation}
    \sigma_3 \approx 1.5 \ 
\end{equation}
since the evidence has a positive singularity for $\sigma = d_1$.
This result is in conflict with the results in Case 1 and Case 2.

\clearpage

\section{BK-inconsistent evidences in trans-\\ dimensional inversion}
\label{Appendix: BK-Inconsist-TransD}
Consider a linear inverse problem
\begin{equation}
\bfd = g(\bfm) = \bfG \bfm 
\end{equation}
with 3 data values $\bfd^{obs} = (d_1, d_2,d_3)$, and two model parameterizations, one with $1$ parameter $\bfm = m$, and the other with $2$ parameters $\bfm = (m_1,m_2$). In the 1-parameter case, $\bfG$ is a single-column matrix: $\bfG_1 = \{ \bfg \}$, and in the 2-parameter case, $\bfG$ is a two-column matrix $\bfG_2 = \{ \bfg^{(1)} \ \bfg^{(2)} \}$. Assume further that the prior is constant in the interval $\calR_1 = [0,{\Delta m}]$ for $1$ parameter, and in the square $\calR_2 = [0,{\Delta m}]\times [0,{\Delta m}]$ for $2$ parameters. If the edge length is ${\Delta m} = 2$, we obtain
\begin{equation}
   p_m(\bfm) = 
   \begin{cases}
       \frac{1}{2} & {\rm for} \ \bfm 
             \in \calR_1 \\
       \frac{1}{4} & {\rm for} \ \bfm 
             \in \calR_2 \\
       0 & \text{otherwise}          
   \end{cases}
\end{equation}
We will represent the data in either Cartesian or spherical coordinates. In the latter case, the relation between the Cartesian data $\bfd = (d_1, d_2,d_3)$ and the spherical data $\breve{\bfd} = (d_r, d_\theta,d_\phi)$ is
\begin{align*}
d_x(d_r, d_\theta,d_\phi) &= d_r \sin d_\theta \cos d_\phi \\
d_y(d_r, d_\theta,d_\phi) &= d_r \sin d_\theta \sin d_\phi \\
d_z(d_r, d_\theta,d_\phi) &= d_r \cos d_\theta
\end{align*}
or, equivalently,
\begin{align}
d_r(d_x, d_y, d_z) &= \sqrt{{d_x}^2 + {d_y}^2 + {d_z}^2} 
\label{eq: Transf-Cart-Sph-1} \\
d_\theta(d_x, d_y, d_z) &= \arccos \frac{{d_z}}{\sqrt{{d_x}^2 + {d_y}^2 + {d_z}^2}}
\label{eq: Transf-Cart-Sph-2} \\
d_\phi(d_x, d_y, d_z) &= \sign(d_y) \arccos \frac{d_x}{\sqrt{{d_x}^2 + {d_y}^2}} \ .
\label{eq: Transf-Cart-Sph-3}
\end{align}
The transformation $S$ from Cartesian to spherical coordinates has the Jacobian 
\begin{equation}
|\det(\bfJ_S)| = \frac{\partial(d_x, d_y, d_z)}{\partial(d_r, d_\theta,d_\phi)} = {d_r}^2 \sin d_\theta \ .
\label{Jac-Cart-Sph}
\end{equation}

\subsection{Case 1: Cartesian data and constant priors}
Assume that $\bfm$ has a uniform prior in a cube $\calC_{\calM}$ with edge length $\Delta m$ in the model space, and that the uncertainties of $\bfd$ are uniformly distributed in Cartesian coordinates, all with the same uncertainty $\sigma$. 
Hence, the data uncertainty distribution is non-zero and constant in a cube $\calC_{\calD}$ in data space, centered at $\bfd^{obs}$, with edge length $2\sigma$. Outside $\calC_{\calD}$ it is zero.

Let us now calculate the evidence for our observations $\bfd^{obs}$ under the assumption that the model parameters are able to fit the data within the error bars. Furthermore, let us for simplicity assume that no model that fits the data has zero prior probability: $g^{-1}(\calC_{\calD}) \subseteq \calC_{\calM}$.

For this calculation we observe that if the number of model parameters $k$ is $2$, the subset of models that fit a single datum $d_i$ ($i = x,y,z$) within error bars $\pm \, \sigma$ lie in a 2-D corridor $g^{-1}( \{d_i \, | \,  d^{obs}-\sigma \le d_i \le d^{obs}+\sigma \}$ between parallel straight lines in parameter space due to the linearity of $g$. For this reason, models that fit all 3 data within their error bars lie in an intersection between 3 such corridors. The result is a polygon in the model space with 3, 4, 5, or 6 sides.

As our example, consider the simple case:
\begin{equation}
\begin{pmatrix}
d_1\\
d_2\\
d_3
\end{pmatrix} =
\begin{Bmatrix}
2 & 1\\
4 & 2\\
1 & 0
\end{Bmatrix} 
\begin{pmatrix}
m_1\\
m_2
\end{pmatrix}
\label{lin-probl-2-A}
\end{equation}

\noindent
where
\begin{equation}
\begin{pmatrix}
d_1\\
d_2\\
d_3
\end{pmatrix} =
\begin{pmatrix}
3.1\\
5.8\\
1.1
\end{pmatrix} \ ,
\end{equation}
are the observed data, and the noise is uniformly distributed with dispersion $\pm \sigma = 0.4$. This means that the models that fit all 3 data within their error bars are located in a parallelogram in model space with corner points: 
\begin{align}
P_1 &= \left(\frac{d_1 - \sigma}{2}, d_3 - \sigma \right)  = 
       \left(1.35, 0.7 \right)\\
P_2 &= \left(\frac{d_2 + \sigma}{4}, d_3 - \sigma \right) = 
       \left( 1.55, 0.7 \right)\\
P_3 &= \left(\frac{d_2 - 2d_3 -\sigma}{4}, d_3 + \sigma \right) = 
       \left( 0.8, 1.5 \right)\\
P_4 &= \left(\frac{d_1 - d_3 - 2\sigma}{2}, d_3 + \sigma \right) = 
       \left( 0.6, 1.5 \right)\ .
\end{align}
The area of this parallelogram is 
\begin{equation}
A  = \frac{\sigma}{2} (d_2 - 2d_1 + 3\sigma) = 0.16 \ .
\end{equation}
The evidence for two model parameters $(k = 2)$ is then (using expression \ref{eq: 2nd likelihood} for the likelihood)
\begin{equation}
p(\bfd_{obs}|2) = \int_\calM p_d(\bfG \bfm) p_m(\bfm|k) d\bfm
\end{equation}
where
\begin{equation}
   p_d(\bfd) = 
   \begin{cases}
       \frac{1}{(2\sigma)^3} & {\rm for} \ \bfd 
             \in \calC_{\calD} \\
       0 & \text{otherwise}          
   \end{cases}
\label{eq: data-distrib-cart-A}
\end{equation}
is the prior in the data space, $\bfd$ is a random variable which contains the measurement noise, and $p_m(\bfm|k)$ is the model prior. Assuming that the prior is nonzero everywhere inside the parallelogram $\calP = P_1 P_2 P_3 P_4$, it contributes with the factor $1/\Delta m^2$ to the evidence, giving
\begin{equation}
p(\bfd_{obs}|2) = \frac{1}{(2\sigma)^3} \, A \, \frac{1}{\Delta m^2} = 
                  \frac{d_2 - 2d_1 + 3\sigma}{(2\sigma)^2 {\Delta m^2}}  \ .
\label{eq: evidence-2D-Cart-A}
\end{equation}
Say we change the number of model parameters to 1, by retaining only $m_1$.  The problem then reads
\begin{equation}
\begin{pmatrix}
d_1\\
d_2\\
d_3
\end{pmatrix} = m_1
\begin{pmatrix}
2\\
4\\
1
\end{pmatrix} \ ,
\end{equation}
where the models that fit all 3 data within their error bars are located on a line segment
\begin{align*}
\calL &= 
\left[ \frac{d_1 - \sigma}{2}, \frac{d_1 + \sigma}{2} \right] \cap
\left[ \frac{d_2 - \sigma}{4}, \frac{d_2 + \sigma}{4} \right] \cap
\left[ \frac{d_3 - \sigma}{1}, \frac{d_3 + \sigma}{1} \right]  \ . \\
      &= \left[ \frac{135}{100},\frac{150}{100} \right]
\end{align*}
 in the $m_1$-space. The length of this line seqment is
\begin{equation}
L = \frac{15}{100} \ .
\end{equation}
The evidence for $k = 1$ therefore becomes:
\begin{equation}
p(\bfd_{obs}|1) = \frac{1}{(2\sigma)^3} \, L \, \frac{1}{\Delta m} \ .
\end{equation}
We can now compute the Bayes factor to be used in the selection of $k$:
\begin{equation}
B = \frac{p(\bfd_{obs}|2)}{p(\bfd_{obs}|1)} = 
\frac{2\sigma (d_2 - 2d_1 + 3\sigma)}{{\Delta m}\, L} 
= 2,13333333\dots
\label{eq: Bayes-fac-Cart-A}
\end{equation}
favoring $k = 2$.


\subsection{Case 2: Spherical data and constant priors}
Let us now reconsider the linear inverse problem in equation \ref{lin-probl-2-A}, but with the difference that the data are now transformed into spherical coordinates via the non-linear transformation (Eqs.~\ref{eq: Transf-Cart-Sph-1}-\ref{eq: Transf-Cart-Sph-3}).

In spherical coordinates, the prior data uncertainties, given by (\ref{eq: data-distrib-cart-A}) in Cartesian coordinates, are now described by the distribution:
\begin{equation}
   p_{S(d)}(d_r,d_\theta,d_\phi) = 
   \begin{cases}
       \frac{1}{(2\sigma)^3}\, d_r^2\, \sin(d_\theta) & {\rm for} \ (d_r,d_\theta,d_\phi)^T 
             \in S(\calC_{\calD}) \\
       0 & \text{otherwise}          
   \end{cases}
\end{equation}
where we have carried out a Jacobian transformation using equation (\ref{Jac-Cart-Sph}). If we replace the data with the forward function, transformed to spherical coordinates, we obtain the likelihood:

{\small
{\flushleft $L_{S(d)} (m_1,m_2) 
  = \frac{1}{(2\sigma)^3}\,  p_{S(d)} (d_r(m_1,m_2), d_\theta(m_1,m_2), d_\phi(m_1,m_2)) =$ }
}
{\small
\begin{align*}
 = \begin{cases}
    \frac{\sqrt{ ( (2 m_1 + m_2)^2 + (4 m_1 + 2 m_2)^2 + m_1^2 ) ( {(2 m_1 + m_2)}^2 + {(4 m_1 + 2 m_2)}^2 )} }{(2\sigma)^3} 
           & {\rm for} \ (m_1, m_2) \in \calP  \\ 
    0      & \text{otherwise}
    \end{cases}
\end{align*}
}To obtain the evidence for $k = 2$, we integrate the product of $L_{S(d)} (m_1,m_2)$ and the prior $1/{\Delta m^2}$ over $\calP$:
\begin{align}
p_{S(d)}(\bfd_{obs}|2) = \frac{1}{\Delta m^2} \int_\calP L_{S(d)} (m_1,m_2)\, dm_1 dm_2 \ .
\label{eq: sph-evid-2-A}
\end{align}
Evaluation of the integral is carried out through the substitution $\bfm = u{\bf a}+v{\bf b} + \bfm_0$ where ${\bf a} = \overline{P_1 P_2}$ and ${\bf b} = \overline{P_1 P_4}$ are vectors spanning the parallelogram $\calP$, and $\bfm_0$ is the location vector of $P_1$. This transformation
\begin{equation}
    \bfm = \bfT(u,v) =          
    \begin{Bmatrix}
           (-1/10+3 \sigma/4) & (-55/100-\sigma/2) \\
           0 & 2 \sigma \\ 
    \end{Bmatrix}
    \begin{pmatrix}
           u \\
           v 
    \end{pmatrix}
\end{equation}
has the Jacobian $J_T = a_1 b_2 - a_2 b_1$ (the area of $\calP$), and we obtain the evidence (see the exact result for the integral over the likelihood in Appendix \ref{Appendix: Exact-evid-sph-transD}):
\begin{align}
p_{S(d)}(\bfd_{obs}|2) &= \int_\calP L_{S(d)} (\bfT(u,v))\, |J_T|\, du\,  dv \\ 
&\approx \frac{5.3682629822 \ (-2 + 15 \sigma) \sigma}{(2\sigma)^3 \Delta m^2}
\label{eq: evidence-2D-Sph} \\
&\approx  \frac{8.58922077}{(2\sigma)^3 \Delta m^2}\ .
\end{align}
where $\sigma = 0.4$, and where we have used expression \ref{eq: 2nd likelihood} for the likelihood. In the case of only one model parameter $m_1$ we obtain the evidence by integrating over the line segment $\calL$ on which the likelihood is non-zero, and multiplying with the prior $1/\Delta m$:
\begin{align}
p_{S(d)}(\bfd_{obs}|1) = \frac{1}{\Delta m} \int_\calL L_{S(d)} (m_1)\, dm_1\ 
\end{align}
where 
{\flushleft $L_{S(d)} (m_1,m_2) = \frac{1}{(2\sigma)^3}\,  p_{S(d)} (d_r(m_1,m_2), d_\theta(m_1,m_2), d_\phi(m_1,m_2))$}
\begin{align*}
  &= \begin{cases}
    \frac{1}{(2\sigma)^3}\, \sqrt{ (2 m_1)^2 + (4 m_1)^2 + m_1^2 ) ( {(2 m_1)}^2 + {(4 m_1)}^2 )} 
           & {\rm for} \ m_1 \in \calL  \\ 
    0      & \text{otherwise}
\end{cases} \\
  &= \begin{cases}
    \frac{1}{(2\sigma)^3}\, 2 \sqrt{105}\, m_1^2
           & {\rm for} \ m_1 \in \calL  \\ 
    0      & \text{otherwise}
\end{cases}
\end{align*}
This gives
\begin{align}
p^S(d_{obs}|1) &= \frac{1}{(2\sigma)^3}\, \frac{1}{\Delta m} \int_\calL 2 \sqrt{105}\ m_1^2\, dm_1\  \\
&= \frac{1}{(2\sigma)^3}\, \frac{1}{\Delta m} \frac{2439}{800}\sqrt{\frac{21}{5}} \approx \frac{6.24807822}{(2\sigma)^3 \Delta m}
\end{align}
From this we obtain the Bayes Factor
\begin{equation}
B = \frac{p^S(\bfd_{obs}|2)}{p^S(\bfd_{obs}|1)} \approx \frac{1}{\Delta m}\frac{8.58922077}{6.24807822} \approx 0,68734901 
\label{eq: Bayes-fac-Sph-A}\ 
\end{equation}
favoring $k = 1$, in contrast to the case with Cartesian data, which favoured $k = 2$.

\clearpage

\section{Expression for the likelihood integral (\ref{eq: sph-evid-2-A})}
\label{Appendix: Exact-evid-sph-transD}
It can be shown that the exact expression for the likelihood integral (\ref{eq: sph-evid-2-A}) is

{\flushleft $\int_\calP L_{S(d)} (m_1,m_2)\, dm_1 dm_2 = $}
{\footnotesize
\begin{align*}
= & \ \frac{20312711127 \sqrt{\frac{409}{5}}-30579261939 \sqrt{\frac{541}{5}}}{14229845000}+\frac{47479907867 \sqrt{\frac{203}{15}}-1620870691 \sqrt{\frac{23849}{5}}}{3484860000} \\
&+\frac{1507 \left( \log \left(21 \sqrt{409}-92 \sqrt{21}\right)-\log \left(21 \sqrt{541}-106 \sqrt{21}\right) \right)}{88200 \sqrt{105}}  \\
& +\frac{8429 \left( \log \left(106 \sqrt{21}+21 \sqrt{541}\right)-\log \left(92 \sqrt{21}+21 \sqrt{409}\right) \right)}{352800 \sqrt{105}} \\
&+\frac{
 31852343043 \left( \sinh ^{-1}\left(\frac{132 \sqrt{5}}{107}\right) -
                    \sinh ^{-1}\left(\frac{1033 \sqrt{5}}{642}\right) \right)}{4646480000 \sqrt{241}}  \\
&+\frac{46582208643 \left( \sinh ^{-1}\left(\frac{1157 \sqrt{5}}{706}\right)
           - \sinh ^{-1}\left(\frac{458 \sqrt{5}}{353}\right) \right)}
           {4646480000 \sqrt{241}}  \\
& + 7 \sqrt{\frac{7}{15}} 
\Bigg(
\frac{\log (7)}{14400}+\frac{\log \left(\frac{7}{3}\right)}{28800}-\frac{\log (1029)}{28800}
+\frac{\log \left(101 \sqrt{3}+3 \sqrt{3407}\right)
-\log \left(21 \sqrt{29}+113\right)}          {7200}  \\
&+\frac{\tanh ^{-1}\left(\frac{113}{21 \sqrt{29}}\right)+
\tanh ^{-1}\left(\frac{92}{\sqrt{8589}}\right)-
\tanh ^{-1}\left(\frac{101}{\sqrt{10221}}\right)-
\tanh ^{-1}\left(\frac{106}{\sqrt{11361}}\right)}{9600}
\Bigg) \ .
\end{align*}
}

\clearpage

\section{Two simple examples showing the acausality of hierarchical Bayes}
\label{Appendix: Simple-Hierarch-B}

\subsection{An example with Gaussian data and model parameters, and \\ discrete-valued hyperparameters}
We will here consider a simple version of the problem, where $\bfd$ and $\bfm$ are scalars ($d$,$m$). We assume that the (partly) unknown priors of $d$ and $m$ are Gaussians with zero mean and unknown standard deviations $\boldsymbol{\theta} = \left(\lambda,\delta \right)^T$:
$$
p_{d|\lambda} = \frac{1}{\lambda \sqrt{2\pi}} \exp\left(-\frac{1}{2} \frac{d^2}{\lambda^2}\right)
$$
$$
p_{m|\delta} = \frac{1}{\delta \sqrt{2\pi}} \exp\left(-\frac{1}{2}\frac{m^2}{\delta^2}\right)
$$
and note that in formulation 2, $\bfd$ is a random variable which contains the measurement noise. Let us assume that the hyper priors on $\boldsymbol{\theta}$ are discrete distributions
$$
p_{\lambda} (\lambda) = 
\begin{cases}
    \pi_\lambda     & \rm{for} \ \lambda = 1  \\ 
    (1-\pi_\lambda) & \rm{for} \ \lambda = 2  \\ 
    0       & \text{otherwise}
\end{cases}
$$
$$
p_{\delta} (\delta) = 
\begin{cases}
    \pi_\delta     & \rm{for} \ \delta = 1  \\ 
    (1-\pi_\delta) & \rm{for} \ \delta = 2  \\ 
    0       & \text{otherwise}
\end{cases}
$$
for $\pi_\lambda \in [0,1]$, $\pi_\delta \in [0,1]$, 
so that the problem is essentially about choosing the ``best'' (most likely) value of each hyper parameter.
Hence, the joint prior is
\begin{equation*}
{\scriptsize
p_{d,m,\lambda,\delta} (d,m,\lambda,\delta) = 
\begin{cases}
    \pi_\lambda \pi_\delta \frac{1}{2\pi} \exp\left(-\frac{1}{2} d^2\right) \exp\left(-\frac{1}{2} m^2\right)  & \rm{for} \ \lambda = 1, \delta = 1  \\ 
    (1-\pi_\lambda) \pi_\delta \frac{1}{4\pi} \exp\left(-\frac{1}{2} \frac{d^2}{4}\right) \exp\left(-\frac{1}{2} m^2\right)  & \rm{for} \ \lambda = 2, \delta = 1 \\ 
    \pi_\lambda (1-\pi_\delta) \frac{1}{4\pi} \exp\left(-\frac{1}{2} d^2\right) \exp\left(-\frac{1}{2} \frac{m^2}{4}\right)  & \rm{for} \ \lambda = 1, \delta = 2 \\
    (1-\pi_\lambda)  (1-\pi_\delta) \frac{1}{8\pi} \exp\left(-\frac{1}{2} \frac{d^2}{4}\right) \exp\left(-\frac{1}{2} \frac{m^2}{4}\right)  & \rm{for} \ \lambda = 2, \delta = 2 \\ 
    0       & \text{otherwise}
\end{cases}}
\end{equation*}
Let us furthermore assume that the problem is linear:
$$
g(m) = k\,m
$$
where $k$ is a constant. Then, using expression \ref{eq: 2nd likelihood} for the likelihood, the posterior becomes:
\begin{flushleft}
$p_{d,m,\lambda,\delta} (g(m),m,\lambda,\delta)/p_d(d) = $
\end{flushleft}
\begin{equation*}
{\footnotesize
\frac{1}{p_d(d_{obs})} \cdot
\begin{cases}
    \pi_\lambda \pi_\delta \frac{1}{2\pi} \exp\left(-\frac{1}{2} \left( k^2 + 1 \right) m^2\right)  & \rm{for} \ \lambda = 1, \delta = 1  \\ 
    (1-\pi_\lambda) \pi_\delta \frac{1}{4\pi} \exp\left(-\frac{1}{2} \left( \frac{k^2}{4} + 1 \right) m^2\right)  & \rm{for} \ \lambda = 2, \delta = 1  \\ 
    \pi_\lambda (1-\pi_\delta) \frac{1}{4\pi} \exp\left(-\frac{1}{2} \left( k^2 + \frac{1}{4} \right) m^2\right)  & \rm{for} \ \lambda = 1, \delta = 2  \\ 
    (1-\pi_\lambda)  (1-\pi_\delta) \frac{1}{8\pi} \exp\left(-\frac{1}{2} \left( \frac{k^2}{4} + \frac{1}{4} \right) m^2\right)  & \rm{for} \ \lambda = 2, \delta = 2  \\ 
    0       & \text{otherwise}
\end{cases}}
\end{equation*}
The discrete marginal posterior for ${\boldsymbol\theta} = (\lambda,\delta)$ is found by integrating over m:
\begin{flushleft}
$p_{\lambda,\delta | d} (\lambda,\delta| d_{obs}) =$
\end{flushleft} 
\begin{equation*}
\frac{1}{p_d(d_{obs})} \cdot
\begin{cases}
    \pi_\lambda \pi_\delta \frac{1}{2\pi} \sqrt{\frac{2\pi}{k^2 + 1}}    & \rm{for} \ \lambda = 1, \delta = 1  \\ 
    (1-\pi_\lambda) \pi_\delta \frac{1}{4\pi} \sqrt{\frac{8\pi}{k^2 + 4}}    & \rm{for} \ \lambda = 2, \delta = 1  \\
    \pi_\lambda (1-\pi_\delta) \frac{1}{4\pi} \sqrt{\frac{8\pi}{4k^2 + 1}}    & \rm{for} \ \lambda = 1, \delta = 2  \\ 
    (1-\pi_\lambda)  (1-\pi_\delta) \frac{1}{8\pi} \sqrt{\frac{8\pi}{k^2 + 1}}  & \rm{for} \ \lambda = 2, \delta = 2
\end{cases}
\end{equation*}
Summation over $\delta$ gives the posterior marginals for the hyper parameters of the data prior:
\begin{equation}
p_{\lambda | d} (\lambda | d_{obs}) = 
\begin{cases}
    -\frac{\pi _{\lambda } \left(\left(\pi _{\delta }-1\right) \sqrt{\frac{1}{4 k^2+1}}-\pi _{\delta
   } \sqrt{\frac{1}{k^2+1}}\right)}{\sqrt{2 \pi }}
    & \rm{for} \ \lambda = 1  \\ 
    \frac{\left(\pi _{\lambda }-1\right) \left(\left(\pi _{\delta }-1\right)
   \sqrt{\frac{1}{k^2+1}}-2 \pi _{\delta } \sqrt{\frac{1}{k^2+4}}\right)}{2 \sqrt{2 \pi }}
  & \rm{for} \ \lambda = 2  
\end{cases}
\label{eq: priori-lambda}
\end{equation}
and summation over $\lambda$ gives the posterior marginals for the hyper parameters of the model prior:
\begin{equation}
p_{\delta | d} (\delta | d_{obs}) =
\begin{cases}
    \frac{\pi _{\delta } \left(\sqrt{\frac{1}{k^2+4}} \left(1-\pi _{\lambda
   }\right)+\sqrt{\frac{1}{k^2+1}} \pi _{\lambda }\right)}{\sqrt{2 \pi }}    & \rm{for} \ \delta = 1  \\ 
    -\frac{\left(\pi _{\delta }-1\right) \left(\sqrt{\frac{1}{k^2+1}} \left(1-\pi _{\lambda
   }\right)+2 \sqrt{\frac{1}{4 k^2+1}} \pi _{\lambda }\right)}{2 \sqrt{2 \pi }}  & \rm{for} \ \delta = 2  
\end{cases}
\label{eq: priori-delta}
\end{equation}
In the solutions (equations \ref{eq: priori-lambda} and \ref{eq: priori-delta}) to this problem we see that the computed prior distributions of $\lambda$ and $\delta$ depend on the forward relation (through $k$), in contradiction with the assumption that $\lambda$ and $\delta$ determine the {\em prior} distributions of noise and model parameters, respectively.

\subsection{A fully Gaussian example of Hierarchical Bayes computation}
We analyze a simple linear example with one datum and one model parameter, both with a Gaussian prior:
\begin{equation}
p_{d|\lambda}(d) = \frac{1}{\lambda \sqrt{2\pi}} \exp\left(-\frac{1}{2} \frac{(d-d_{obs})^2}{\lambda^2}\right)
\end{equation}
\begin{equation}
p_{m|\delta}(m) = \frac{1}{\delta \sqrt{2\pi}} \exp\left(-\frac{1}{2}\frac{(m-m_0)^2}{\delta^2}\right)
\end{equation}

\bigskip\noindent
We assume that the hyper priors $\boldsymbol{\theta} = [\lambda, \delta]^T$ are Gaussian with zero mean and known standard deviations $\sigma_\lambda$ and $\sigma_\delta$, respectively:
$$
p_{\lambda}(\lambda) = \frac{1}{\sigma_\lambda \sqrt{2\pi}} \exp\left(-\frac{1}{2} \frac{\lambda^2}{{\sigma_\lambda}^2}\right)
$$
$$
p_{\delta}(\delta) = \frac{1}{\sigma_\delta \sqrt{2\pi}} \exp\left(-\frac{1}{2} \frac{\delta^2}{{\sigma_\delta}^2}\right).
$$
The joint prior is
{\flushleft 
$\frac{1}{A}\ p_{d,m,\lambda,\delta} (d,m,\lambda,\delta) = $}
\begin{align*}
&= 
\frac{1}{A} \cdot p_{d} (d) p_{m} (m) 
p_{\lambda}(\lambda)
p_{\delta}(\delta) \\
&=\frac{1}{A \cdot 4\pi^2 \lambda \delta\, {\sigma_\lambda} {\sigma_\delta}}
\exp\left(
-\frac{1}{2} 
\left( 
\frac{(d-d_{obs})^2}{\lambda^2} +
\frac{(m-m_0)^2}{\delta^2} +
\frac{\lambda^2}{\sigma_\lambda^2} +
\frac{\delta^2}{\sigma_\delta^2}
\right)
\right) 
\end{align*}
where $A$ is a normalization constant. Assuming that the priors of $d$ and $m$ have means $d_{obs} = 1$ and $m_0 = 1$ and unknown standard deviations $\boldsymbol{\theta} = \left(\lambda,\delta \right)^T$, respectively, and that $\sigma_\lambda = 1$ and $\sigma_\delta = 1$, we obtain
{\small
\begin{equation*}
\frac{1}{A}\ p_{d,m,\lambda,\delta} (d,m,\lambda,\delta) = 
\frac{1}{A \cdot 4\pi^2 \lambda \delta }
\exp\left(
-\frac{1}{2} 
\left( 
\frac{(d-1)^2}{\lambda^2} +
\frac{(m-1)^2}{\delta^2} +
\lambda^2 +
\delta^2
\right)
\right)
\end{equation*}}Inserting the forward function $d = k\cdot m$ (see equation \ref{eq: Bayes-2}), and integrating over $m$, we obtain the posterior over $(\lambda,\delta)$:
\begin{figure}[h!]
\begin{center}
\includegraphics[width=3.5in]{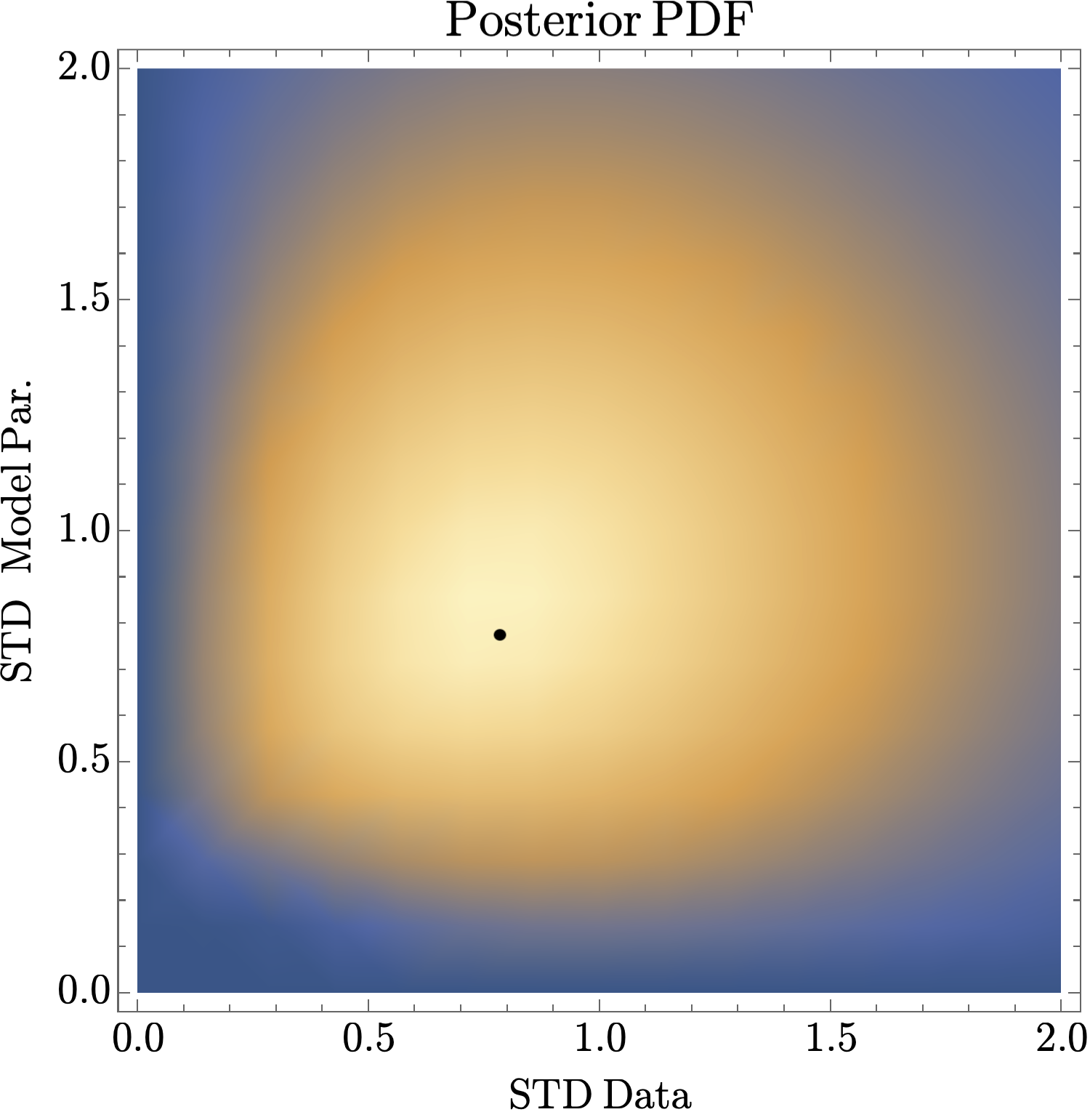}
\caption{\small Calculated joint posterior probability density for $(\lambda,\sigma)$, the `prior' standard deviations of data and model parameter, respectively. The black dot indicates the location of the maximum posterior point.}
\label{joint-lambda-sigma-post}
\end{center}
\end{figure}

{\flushleft 
$\frac{1}{A'}\ p_{\lambda,\delta} (\lambda,\delta) = 
\int \frac{1}{A}\ p_{d,m,\lambda,\delta} (km,m,\lambda,\delta)\ dm = $}
{\small
\begin{align}
&=\frac{1}{A \cdot 4\pi^2 \lambda \delta }
\nonumber
\exp\left(
-\frac{1}{2} 
\left(
\lambda^2 +
\delta^2
\right)
\right)
\int \exp\left(
-\frac{1}{2} 
\left( 
\frac{(km-1)^2}{\lambda^2} +
\frac{(m-1)^2}{\delta^2}
\right)
\right) dm \\ 
&=\frac{1}{A \cdot 4\pi^2 \lambda \delta } \cdot  
\exp\left(
-\frac{1}{2} 
\left(
\lambda^2 +
\delta^2
\right)
\right)
\sqrt{\frac{2 \pi}{\frac{k^2}{\lambda^2}+\frac{1}{\delta^2}}} \exp\left({-\frac{(k-1)^2}{2 \left(\lambda^2+\delta^2 k^2\right)}}\right) \ .
\label{posterior-vars}
\end{align}
}
Ignoring for a moment the BR-inconsistency in MAP estimates described in Appendix A, we can now choose the standard deviations $\lambda,\sigma$ that maximize this posterior (see figure \ref{joint-lambda-sigma-post}), but
we see that the computed prior distributions of $\lambda$ and $\delta$ vary with the forward relation (through $k$), in contradiction with the claim that $\lambda$ and $\delta$ determine the {\em prior} distributions of noise and model parameters, respectively.

\clearpage

\section{Non-uniqueness of evidences in overdetermined inverse problems}

\label{Appendix: Non-uniqueness}
\subsection{Finding a probability density over an N-dimensional cube with a given integral over a given sub-manifold }
\label{subsubsec: density-w-given-evidence}
Let us first construct a probability density function $ f(\mathbf{x}) $ on the $ N $-dimensional domain $ \calD = [0, 1]^N $, where $ \mathbf{x} = (\mathbf{x}_1,\mathbf{x}_2)  $, $ \mathbf{x}_1 \in [0, 1]^k $ and $ \mathbf{x}_2 \in [0, 1]^{N-k} $, such that it has a constant value $ c $ everywhere on the submanifold $ \calS = \{(\mathbf{x}_1, g(\mathbf{x}_1)) \mid \mathbf{x}_1 \in [0,1]^k\} $, where $ g(\mathbf{x}_1) $ is a differentiable function.

The suggested form for $ f(\mathbf{x}) $ is:

\[
f(\mathbf{x}) = \frac{A}{\sqrt{(2\pi)^{N-k} \sigma^{2(N-k)}}} \exp\left( - \frac{\|\mathbf{x}_2 - g(\mathbf{x}_1)\|^2}{2 \sigma^2} \right)
\]

where:
\begin{itemize}
    \item $ A $ is a normalization constant, ensuring that the integral over the domain $ \calD $ equals 1,
    \item $ \sigma $ is a parameter controlling the width of the density around the submanifold,
    \item $ \|\mathbf{x}_2 - g(\mathbf{x}_1)\|^2 $ is the squared Euclidean distance between $ \mathbf{x}_2 $ and the submanifold $ \mathbf{x}_2 = g(\mathbf{x}_1) $.
\end{itemize}
This density has the desired property of being concentrated around the submanifold $ \mathbf{x}_2 = g(\mathbf{x}_1) $ and is normalized on $ \calD $.

\bigskip\noindent
If we want that $ f(\mathbf{x}) = c $ \ everywhere on $ \calS $, we can choose $ \sigma $ such that 
\[
\frac{A}{\sqrt{(2\pi)^{N-k} \sigma^{2(N-k)}}} = c \ ,
\]
or,
\[
\sigma = \left( \frac{1}{(2\pi)^{N-k}} \left( \frac{A}{c} \right)^2 \right)^{\frac{1}{2(N-k)}} \ .
\]
If $ V $ is the volume/area/length of the submanifold $ \calS $ within $ \calD $, the integral of $ f(\mathbf{x} | \calS)$ will be $ c\, V $, and this can be adjusted to any positive number by choosing an appropriate value of $ \sigma $.

\bigskip\noindent
The above result shows that, for an overdetermined inverse problem where the image of the parameter space under the forward mapping is a sub-manifold $ \calS $ of the data space $ \calD $, and for any positive number $ E $, there exist a probability density over $ \calD $, such that its integral over $ \calS $ is equal to $ E $. In other words, any evidence can be obtained by choosing an appropriate $ \sigma $ in $ f(\mathbf{x}) $.

\bigskip\noindent
To complete our demonstration of the non-uniqueness of evidences in inverse problems, we will now show that, for any two different probability densities over $\calD$, there will always exist a re-parameterization of $\calD$ that transforms one density into the other.

\subsection{A transformation between two given probability densities}
Given a probability density function $ f(\mathbf{x}) $ on the $ N $-dimensional domain $ [0, 1]^N $, we seek a transformation $ T(\mathbf{x}) = \mathbf{u} $ that transforms $ f(\mathbf{x}) $ into a desired probability density $ g(\mathbf{u}) $. The relationship between $ f(\mathbf{x}) $ and $ g(\mathbf{u}) $ is given by:
\[
f(\mathbf{x}) = g(\mathbf{u}) \left| \det(J) \right|
\]
where $ J  =  \partial \mathbf{u}/\partial \mathbf{x} $ is the Jacobian matrix of the transformation $ T(\mathbf{x}) = \mathbf{u} $. The transformation can also be expressed through the joint cumulative distribution functions
   \[
   F_f(\mathbf{x}) = \int_0^{x_1} \cdots \int_0^{x_N} f(x_1', \dots, x_N') \, dx_N' \dots dx_1'
   \]
and
   \[
   F_g(\mathbf{u}) = \int_0^{u_1} \cdots \int_0^{u_N} g(u_1', \dots, u_N') \, du_N' \dots du_1'
   \]
showing that the transformation must satisfy the simple relation
\[
F_f(\mathbf{x}) = F_g(\mathbf{u}) \ .
\]
For each coordinate $ x_k $, we now transform it into $ u_k $ using the marginal and conditional CDFs:
\renewcommand{\theenumi}{\alph{enumi}}
\begin{enumerate}
\item First Coordinate:
   \[
   u_1 = F_1(x_1)
   \]

\item Second Coordinate:
   \[
   u_2 = F_2(x_2 | x_1)
   \]

\item Subsequent Coordinates:
   \[
   u_k = F_k(x_k | x_1, \dots, x_{k-1})
   \]
\end{enumerate}
This constructs the transformation $ T(\mathbf{x}) = \mathbf{u} $ \ that transforms $ f(\mathbf{x}) $ into $ g(\mathbf{u}) $.

\bigskip\noindent
This result, together with the result in \ref{subsubsec: density-w-given-evidence}, shows that, for an overdetermined inverse problem with parameter space $ [0, 1]^N $, any evidence can be obtained by choosing an appropriate re-parameterization in the data space. This result remains valid for any overdetermined inverse problem with parameters defined in a general $ N $-dimensional interval, since the domain of each parameter can be transformed into $ [0, 1] $ through simple linear transformations.

\clearpage

\section{Standard derivation of the likelihood in Appendix \ref{Appendix: BK-incon-Hierarch}, Case 2: Transformed data with $d_1 \rightarrow \tan(d_1)$.} 

\label{Appendix: Likelihood-Formulation-1}
We reconsider the linear inverse problem $\bfd^{obs} = \bfG \bfm + \hat{\bfn}$: 
\begin{equation}
\begin{pmatrix}
d_1\\
d_2\\
d_3
\end{pmatrix} =
\begin{Bmatrix}
0 & a\\
b & 0\\
c & 0
\end{Bmatrix} 
\begin{pmatrix}
m_1\\
m_2
\end{pmatrix}
+ \begin{pmatrix}
\hat{n}_1\\
\hat{n}_2\\
\hat{n}_3
\end{pmatrix}
\label{lin-probl-2}
\end{equation}
where $\bfd^{obs} = (d_1, d_2, d_3)^T$ is the observed data, and $\bfm = (m_1, m_2)^T$ is the unknown parameter vector. The noise $\hat{\bfn} = (\hat{n}_1, \hat{n}_2, \hat{n}_3)^T$ has the distribution
\begin{equation}
   p_d(\hat{\bfn}) = 
   \begin{cases}
       \frac{1}{(2\sigma)^3} & {\rm for} \ \hat{\bfn} 
             \in [-\sigma,\sigma]^3 \\
       0 & \text{otherwise} \ .
   \end{cases} 
\label{eq: data-distrib-cart}
\end{equation}
Since this distribution is symmetric around the origin, the sign-reversed noise (to be used below) $\bfn = -\hat{\bfn}$ has the same distribution:
\begin{equation}
   p_d(\bfn) = 
   \begin{cases}
       \frac{1}{(2\sigma)^3} & {\rm for} \ \bfn 
             \in [-\sigma,\sigma]^3 \\
       0 & \text{otherwise} \ .      
   \end{cases} 
\label{eq: data-distrib-cart}
\end{equation}
Data and uncertainties are now reparameterized via the transformation
\begin{equation}
\begin{pmatrix}
y_1\\
y_2\\
y_3
\end{pmatrix} =
T_1
\begin{pmatrix}
d_1\\
d_2\\
d_3
\end{pmatrix} =
\begin{pmatrix}
\tan(d_1)\\
d_2\\
d_3
\end{pmatrix} \ ,
\end{equation}
with inverse Jacobian
\begin{equation}
|\det(\bfJ_S)| = \left| \frac{\partial(d_1, d_2, d_3)}{\partial(y_1, y_2, y_3)} \right| = \frac{1}{y_1^{2}+1} \ .
\label{Jac-arctan}
\end{equation}
The transformed forward function is
\begin{equation}
    g_y(\bfm) = 
    \begin{pmatrix}
        \tan(a m_2)\\
        bm_1\\
        cm_1
   \end{pmatrix} \ .
\end{equation}
Substituting $y_1 = y_1^{obs}+\hat{n}_1 = y_1^{obs}-n_1$ into equation \ref{Jac-arctan}, the Jacobian transformation provides the distribution of noise on the transformed data $\bfy$:
\begin{equation*}
    p_n^{y} (\bfn) = 
       \begin{cases}
       \frac{1}{(2\sigma)^3}\frac{1}{(y_1^{obs}-n_1)^{2}+1} & {\rm for} \ \bfn \in 
          [\tan(d_1-\sigma)-\tan(d_1),\tan(d_1+\sigma)-\tan(d_1)] \times [-\sigma,\sigma]^2 \\
       0 & \text{otherwise}          
   \end{cases}
\end{equation*}
where we have carried out a Jacobian transformation using equation (\ref{Jac-arctan}). 

The standard likelihood expression for the inverse problem with $\bfy$ as data is (see our {\em Formulation 1} or, e.g., Calvetti and Somersalo, 2017):
\begin{align*}
    L_y(\bfm) &= p_n^{y} (\bfy - g_y(\bfm)) = 
        p_n^{y} \left(
        \begin{pmatrix}
            y_1 - \tan(a m_2)\\
            y_2 - bm_1\\
            y_3 - cm_1
        \end{pmatrix} \right) \\
        &= 
         \begin{cases}
            \frac{1}{(2\sigma)^3}\frac{1}{\tan(a m_2)^{2}+1} & {\rm for} \ \bfm \in \calP(\sigma)  \\
            0 & \text{otherwise}
         \end{cases} \\
         &= 
         \begin{cases}
            \frac{1}{(2\sigma)^3}\cos^2(a m_2) & {\rm for} \ \bfm \in \calP(\sigma)  \\
            0 & \text{otherwise}
         \end{cases}
\end{align*}
where $\calP(\sigma)$ is the domain in the parameter space with non-zero likelihoods. The above result is identical with the solution we obtained in Appendix \ref{Appendix: BK-incon-Hierarch} with formulation 2 of Bayes Formula.

\end{appendices}

\clearpage

\end{document}